\documentclass{aa}

\usepackage{natbib}
\usepackage{graphicx}
\usepackage{amssymb}
\usepackage{color}

\newcommand{\bl}[1]{\mbox{\boldmath$ #1 $}}

\begin{document}

\title{Stellar hydrodynamical modeling of dwarf galaxies: simulation methodology, tests, and first
results}

\author{Eduard I. Vorobyov\inst{1,2}, Simone Recchi\inst{1}, and Gerhard Hensler\inst{1}}

\institute{Institute for Astrophysics, University of Vienna, T\"urkenschanzstrasse 17, Vienna 1180, Austria \and 
Research Institute of Physics, Southern Federal University, Stachki 194, Rostov-on-Don, Russia}

\abstract {In spite of enormous progress and brilliant
  achievements in cosmological simulations, they still lack
  numerical resolution or physical processes to simulate 
  dwarf galaxies in sufficient details.  Accurate numerical simulations of 
  individual dwarf galaxies are thus still in demand.}  
  {We aim at improving available
  numerical techniques to simulate individual dwarf galaxies.  In particular,
  we aim at $(i)$ studying in detail the coupling between stars and
  gas in a galaxy, exploiting the so-called stellar hydrodynamical
  approach, and $(ii)$ studying for the first time the chemo-dynamical
  evolution of individual galaxies starting  from self-consistently calculated
  initial gas distributions.} 
  {We present a novel chemo-dynamical code for studying the evolution of individual dwarf
  galaxies.   In this code, the dynamics of gas is computed using the usual hydrodynamics 
  equations, while the dynamics of stars is described by the stellar hydrodynamics
  approach, which solves for the first three moments of the collisionless Boltzmann equation.  The
  feedback from stellar winds and dying stars is followed in detail.
  In particular, a novel and detailed approach has been developed to
  trace the aging of various stellar populations, which enables an accurate 
  calculation of the stellar feedback depending on the stellar age. 
  The code has been accurately benchmarked, allowing us to provide a recipe for improving the 
  code performance on the Sedov test problem. } 
  {We build initial equilibrium models of dwarf galaxies that take gas self-gravity into account and
  present different levels of rotational support. Models with high
  rotational support (and hence high degrees of flattening) develop
  prominent bipolar outflows; a newly-born stellar population in these 
  models is preferentially concentrated to the galactic midplane.  
  Models with little rotational support blow 
  away a large fraction of the gas and the resulting stellar distribution is 
  extended and diffuse.   Models that start from non-self-gravitating initial
  equilibrium configurations, evolve at a much slower pace owing to lower initial
  gas density and hence lower star formation rate.
  The stellar dynamics turns  out to be a crucial aspect of galaxy evolution.  
  If we artificially suppress stellar dynamics, supernova explosions occur in a medium
  heated and diluted by the previous activity of stellar winds, thus 
  artificially enhancing the stellar feedback.}
 {The stellar hydrodynamics approach presents a promising tool for studying numerically 
 the coupled evolution of gas and stars in dwarf galaxies.}

\authorrunning{Vorobyov et al.}
\titlerunning{Stellar hydrodynamics and evolution of dwarf galaxies}
\maketitle

\section{Introduction}
Cosmological simulations are reaching nowadays extraordinary levels of
complexity and sophistication.  Based on the so-called standard model
of cosmology (SMoC), these state-of-the-art simulations are able to
reproduce the main features of observed large galaxies \citep[see
e.g.][]{brook12, zemp12, hopk13, voge14, perret14, schaye14}, although
severe disagreements with observations still persist \citep[see
e.g.][]{genel14}.  However, the resolution of these simulations still
does not allow to treat small-scale physics in dwarf galaxies (DGs).  
  
Unfortunately, most of the weaknesses and problems shown by the SMoC
manifest themselves in the range of masses typical for DGs \citep[see
e.g.][for recent reviews]{kroupa12, fm12, fm13}.  Besides full-blown
cosmological simulations, it is thus necessary to run simulations of
individual galaxies or small groups of galaxies.  This enables
studying physical processes in more numerical resolution and
address the problems of the SMoC.  The simulation of individual
galaxies is therefore still extremely relevant.  Also in this field
very remarkable level of accuracy have been reached \citep[see
e.g.][among others]{hopk13, rena13, roskar13, minchev13, teys13}.  The
simulation of individual galaxies can also be used to parameterise
feedback effects that can not be treated in detail in cosmological
simulations \citep{crea13, recchi14}.

An obvious drawback of simulating individual galaxies is that it is
not clear what initial conditions one should adopt.  A common strategy
is to consider a rotating, isothermal gas in equilibrium with the
potential generated by a fixed distribution of stars and/or of dark
matter (DM), but not with the potential generated by the gas itself
\citep{ti88, stt98, mf99, rh13}.  A typical justification for
neglecting gas self-gravity in DG simulations is that the mass budget
of these objects should be dominated by dark matter.  In our previous
paper \citep{VRH1}, we have explained in detail why this approach
may produce unrealistic initial gas distributions and considered
also the gas self-gravity when building equilibrium initial conditions.

As known from stellar structure studies \citep[see e.g.][]{om68}, the
inclusion of gas self-gravity leads to implicit equations describing
the initial equilibrium configuration, which must be solved
iteratively.  We have done that for different halo masses, different
degrees of rotational support and different initial temperatures and
we have calculated the amount of gas prone to star formation for each
of these models.  Our star formation criteria were based on the Toomre
instability criterium and on the Kennicutt-Schmidt empirical
correlation between gas surface density and star formation rate (SFR)
per unit area \citep[see][for more details]{VRH1}.  For models that
satisfy both star formation criteria, the underlying assumption was
that the initial configuration is marginally stable to star formation
and an external perturbing agent can trigger star formation in the
corresponding parts of the galaxy.

The subsequent logical step is to use these marginally stable
configurations as initial conditions for full-blown hydrodynamical
simulations.  For this purpose, we have developed a detailed 2+1
dimensional hydrodynamical code in cylindrical coordinates with
assumed axial symmetry, which treats both the stellar and gaseous
components as well as the phase transitions between them.  Of course,
a real galaxy may depart from such axial symmetry.
However, since our initial conditions are axially symmetric and we do
not explicitly consider environmental effects in this paper, 
 the departures from axial
symmetry are not expected to be significant.

This is the first of a series of papers devoted to numerically
studying DGs with self-consistent initial conditions.  Here, we
explain in detail the basic strategy and the employed numerical
hydrodynamical methods, which we extensively benchmark.  Furthermore,
we run a few representative models to show the overall early evolution
of a typical DG.  We pay particular attention to the co-evolution of
stars and gas in our simulations.  We follow the motion of stars by
means of the so-called stellar hydrodynamical approach \citep{Theis92,Samland,VT06,Mitchell,Kulikov2014}, 
according to which the stellar component of a galaxy can
be treated as a fluid. We develop this approach further and describe
how stars and gas can be coupled by means of star formation and
stellar feedback processes.

The paper is organised as follows. In section \ref{methods} we
describe the simulation methodology, paying specific attention to
coupling the stars and gas via star formation and stellar feedback.
Section \ref{coolheat} contains the gas cooling and heating rates
employed in our modeling. The solution method of our stellar
hydrodynamics equations is described in Section~\ref{solution}.  The
evolution of several representative models of DGs is studied in
Section~\ref{sec:results}. Model caveats are discussed in 
Section~\ref{caveats}. The main conclusions are summarized in
Section~\ref{conclusions} and we benchmark our code against a suite of
test problems in the Appendix.

\section{Simulation methodology}
\label{methods}
Our model DGs consist of a gas disk, stellar component, and
fixed dark matter halo.  The gas disk is a mixture of nine chemical
elements (H, He, C, O, N, Ne, Mg, Si, and Fe), with the initial
abundance of heavy elements set by the adopted initial metallicity.
The stellar component consists of new stars born in the course of
numerical simulations and has no pre-existing component.  The DM halo
has a spherically symmetric form and contributes only to the total
gravitational potential of the system.

We describe the time evolution of our model galaxies using the coupled
system of gas and stellar hydrodynamics equations complemented with
phase transformations between stars and gas, chemical enrichment by
supernova explosions, low- and intermediate-mass stars and also star
formation feedback.  We assume that individual chemical elements are
well coupled dynamically with the bulk motion of the gas, which means
that we need to solve only for the continuity equation for every
chemical element with mass density $\rho_i$ and need not to solve for
the dynamics of every element separately. Below, we provide a brief
explanation for the key equations used to evolve our system in time.

\subsection{Gas hydrodynamics}
The dynamics of gas is modelled using the usual hydrodynamics
equations for the gas density of $\rho_{\rm g}$, momentum $\rho_{\rm
  g} v_{\rm g}$, and internal energy density $\epsilon$ complemented
with the rates of mass, momentum and energy transfer between the gas
and stellar components and also with the continuity equations for the
mass density of every chemical element $\rho_{\rm i}$
\begin{equation}
{\partial \rho_{\rm g} \over \partial t} + {\bl \nabla} \cdot \left(\rho_{\rm g} {\bl v}_{\rm g} \right)
= {\cal S} - {\cal D},
\label{contingas}
\end{equation}
\begin{equation}
{\partial \rho_{\rm i} \over \partial t} + {\bl \nabla} \cdot \left(\rho_{\rm i} {\bl v}_{\rm g} \right)
= {\cal S}^{i} - {\cal D}^{i},
\end{equation}
\begin{eqnarray}
{\partial \over \partial t} \left (\rho_{\rm g} {\bl v}_{\rm g} \right)  
+ {\bl \nabla } \cdot \left ( \rho_{\rm g} {\bl v}_{\rm g} \otimes  {\bl v}_{\rm g}  \right) &=&
-{\bl \nabla} P +\rho_{\rm g} {\bl g}_{\rm gr} +  \nonumber \\
  &+&  {\bl v}_{\rm s}{\cal S} - {\bl v}_{\rm g}{\cal D},
\label{momgas}  
\end{eqnarray}
\begin{equation}
{\partial \epsilon \over \partial t} + {\bl \nabla} \cdot \left( \epsilon {\bl v}_{\rm g} \right) 
= - P {\bl \nabla} \cdot {\bl v}_{\rm g} -\Lambda + \Gamma + \Gamma_\ast - \epsilon {{\cal D}
\over \rho_{\rm g}}.
\label{energy}
\end{equation}
Here, ${\bl v}_{\rm g}$ and ${\bl v}_{\rm s}$ are the gas and stellar
velocities, $P=(\gamma-1)\epsilon$ is the gas pressure linked to the
internal energy density via the ideal equation of state with the ratio
of specific heats $\gamma=5/3$ and ${\bl g}_{\rm gr}$ is the
gravitational acceleration due to gas, stars, and DM halo.  The gas
cooling $\Lambda$ due to line and continuum emission and gas heating
$\Gamma$ due to cosmic rays and small dust grains are explained in
detail in Section~\ref{coolheat}.  The gas heating due to stellar
feedback $\Gamma_\ast$ is calculated as the sum of the contributions
from SNII and SNIa explosions ($\Gamma_{\rm SNII}$ and $\Gamma_{\rm
  SNIa}$, respectively), and also from the stellar winds $\Gamma_{\rm
  sw}$.  The last term on the r.h.s. of equation~(\ref{energy})
accounts for the decrease in the internal energy due to formation of
stars from the gas phase.  We note that the quantity $\rho_{\rm g}
{\bl v}_{\rm g} \otimes {\bl v}_{\rm g}$ is the symmetric dyadic (a
rank-two tensor) calculated according to the usual rules and ${\bl
  \nabla } \cdot \left ( \rho_{\rm g} {\bl v}_{\rm g} \otimes {\bl
    v}_{\rm g} \right)$ is the divergence of a rank-two tensor.


The source term $\cal S$ denotes the phase transformation of stars
into the gas phase and is defined as the sum of the mass release rates
(per unit volume) of all individual elements, ${\cal S}=\sum_{i} {\cal
  S}_{\rm i}$.  The mass release rate of a particular element $i$ is
defined as
\begin{equation}
{\cal S}^{i}={\cal S}^{i}_{\rm SNII} +{\cal S}^{i}_{\rm SNIa} + {\cal S}^{i}_\ast,
\end{equation}
where ${\cal S}^{i}_{\rm SNII}$, ${\cal S}^{i}_{\rm SNIa}$, and ${\cal
  S}^{i}_\ast$ are the contributions due to SNII, SNIa, and also due
to low- and intermediate mass stars, respectively. The sink term $\cal
D$ describes the phase transformation of gas into stars according to
the adopted star formation law.  The detailed explanation of all
relevant rates is provided in Section~\ref{SSterms}.

\subsection{Stellar hydrodynamics}
The time evolution of the stellar component is computed using the
Boltzmann moment equation approach 
introduced by \citet{Burkert88}and further developed in application to stellar
disks in \citet{Samland} and \citet{VT06}. In this approach, the
dynamics of stars is described by the first three moments of the
collisionless Boltzmann moment equations
\begin{equation}
{\partial \rho_{\rm s} \over \partial t} + {\bl \nabla} \cdot \left(\rho_{\rm s} {\bl v}_{\rm s} \right)
= {\cal D} - {\cal S},
\label{continstar}
\end{equation}
\begin{eqnarray}
{\partial \over \partial t} \left (\rho_{\rm s} {\bl v}_{\rm s} \right)  
+ {\bl \nabla } \cdot \left ({\bl v}_{\rm s} \otimes \rho_{\rm s}  {\bl v}_{\rm s}  \right) &=&
-{\bl \nabla} \cdot {\bl\Pi} +\rho_{\rm s} g_{\rm gr} - \nonumber \\
  &-&  {\bl v}_{\rm s}{\cal S} + {\bl v}_{\rm g}{\cal D},
\label{momstar}  
\end{eqnarray}
\begin{eqnarray}
{\partial \Pi_{ij} \over \partial t} + {\bl \nabla} \cdot \left( \Pi_{ij} {\bl v}_{\rm s} \right) &=&
- \Pi_{ik}:({\bl \nabla}{\bl v}_{\rm s})_{jk} - \Pi_{jk}:({\bl \nabla}{\bl v}_{\rm s})_{ik} 
- \nonumber \\
 &-& \sigma_{ij}^2 {\cal S} + \sigma_{\ast}^2 {\cal D}, 
\label{dispstar}
\end{eqnarray}
where $\rho_{\rm s} = m \int f d^3{\bl u}$ is the stellar mass density,
$\bl{v}_{\rm s}=\rho_{\rm s}^{-1} m \int f \,{\bl u}\, d^3{\bl u}$ is
the mean stellar velocity, and $\bl \Pi$ is the stress tensor with 
six components $\Pi_{ij}\equiv\rho_{\rm s}
\sigma_{ij}^2$. The stellar velocity dispersions are defined as
$\sigma^2_{ij}=\rho^{-1} m \int f\, (u_i-v_i)(u_j-v_j)\,d^3{\bl u}$.
The function $f$ is the distribution function of stars in the six
dimensional position-velocity phase space $f\equiv f(t,{\bl x},{\bl
  u})$ and $m$ is the average mass of a star.  The quantity ${\bl
  \nabla} {\bl v}_{\rm s}$ is the gradient of a vector, the covariant
expression for which is provided, e.g., in \citet{SN92} and
$\Pi_{ik}:({\bl\nabla}{\bl v}_{\rm s})_{jk}$ is the convolution (over
index $k$) of two rank-two tensors.  The quantity $\sigma_\ast$
represents a typical stellar velocity dispersion of a new-born cluster
of stars, for which we take a fiducial value of 5.0~km~s$^{-1}$.

\subsection{Massive stars}
Massive stars that end up their life with SNII explosions are main
contributors to the energy budget and chemical composition of the
interstellar medium. Therefore, it is important to follow their
evolution as accurate as possible.
We improve the description of the stellar component by solving for a
separate continuity equation for the mass volume density $\rho^{\rm
  h}_{\rm s}$ of stars with mass $> 8.0~M_\odot$
\begin{equation}
{\partial \rho^{\rm h}_{\rm s} \over \partial t} + {\bl \nabla} \cdot \left(\rho^{\rm h}_{\rm s} 
{\bl v}_{\rm s} \right) = f_{\rm h}{\cal D} - {\cal S^{\rm h}},
\label{continstar_m}
\end{equation}
where $f_{\rm h}$ is the fraction of stars with mass $>8.0~M_\odot$
calculated according to the adopted IMF (see equation~\ref{function})
and ${\cal S}_{\rm h}$ is the rate of death of massive stars.  The
massive star subcomponent has the same dynamical properties as the
rest of the stars, an assumption that we plan to relax in the future.

\subsection{Intermediate-mass stars}
Stars in the $(1.0-8.0)~M_\odot$ mass range produce the majority
  of carbon and nitrogen.  Moreover, binary stars in this mass range
can explode as Type~Ia supernovae and are thus important sources of
iron.  Therefore, we also solve for a separate continuity equation for
the mass volume density $\rho^{\rm m}_{\rm s}$ of intermediate-mass
stars
\begin{equation}
{\partial \rho^{\rm m}_{\rm s} \over \partial t} + {\bl \nabla} \cdot 
\left(\rho^{\rm m}_{\rm s} 
{\bl v}_{\rm s} \right) = f_{\rm m}{\cal D} - {\cal S^{\rm m}},
\label{continstar_i}
\end{equation}
where $f_{\rm m}$ is the fraction of stars with mass
$1.0~M_\odot<m<8.0~M_\odot$ calculated according to the adopted IMF
and ${\cal S}_{\rm m}$ is the rate of death of intermediate-mass
stars. We note that the stellar evolution of low-mass stars can be neglected
during the galactic evolution. They only contribute to the gravitational
potential where their density can be deduced from $\rho_{\rm s}$ 
(the total stellar density), $\rho_{\rm s}^{\rm h}$, and $\rho_{\rm s}^{\rm m}$.

\subsection{Gravity of gas, stars, and DM halo}
The gravitational potential of gas and stars is calculated
self-consistently by solving for the Poisson equation
\begin{equation}
{\bl \nabla}^2 \Phi_{\rm g+s}= 4\pi G \left(\rho_{\rm g}+\rho_{\rm s} \right).
\label{poisson}
\end{equation}
To accelerate the simulations, we solve for Equation~(\ref{poisson})
only when the relative error in the currently stored gravitational
potential compared to the current density distribution exceeds
$10^{-5}$ \citep[see][for details]{SN92}.

The gravitational acceleration due to a spherically symmetric DM halo
${\bl g}_{\rm h}$ is calculated as explained in \citet{VRH1}. Two
options are available in the code: a cored isothermal DM
halo profile and a Navarro-Frenk-White (or cuspy) profile.  In this
paper, we use the cored isothermal DM profile. The total gravitational
acceleration is calculated as ${\bl g}_{\rm gr}={\bl g}_{\rm h}-{\bl
  \nabla} \Phi_{\rm g+s}$.

\subsection{Mean stellar age}
\label{meanstarage}
In order to calculate the mass and energy release rates by supernovae
and dying low-mass stars one needs to know the age of stars. In the
stellar hydrodynamics approach, stars are a mixture of populations
with different ages. Nevertheless, one may define the mean age of stars
${\overline\tau}_\ast$, which should obey the following rules: 
\begin{enumerate} 
\item The bulk motion of stars, such as compression or rarefaction,
  should not affect the value of ${\overline\tau}_\ast$. This can be
  achieved by solving for the Lagrangian or comoving equation for ${\overline\tau}_\ast$
\begin{equation}
\label{advectTau}
{d {\overline\tau}_\ast \over d t} \equiv { \partial {\overline\tau}_\ast \over \partial t} +
({\bl v}_{\rm s} \cdot {\bl \nabla }) {\overline\tau}_\ast =0 
\end{equation}  
\item A newly-born stellar population should rejuvenate the
  pre-existing one. This is achieved by updating the mean age every timestep in every grid cell 
  using the following equation
\begin{equation}  
\label{rejuv}
{\overline\tau}_\ast^{\rm rjv}={ M_{\rm s} {\overline\tau}_\ast +
\triangle M_{\rm s} \, \triangle t \over M_{\rm s} +\triangle M_{\rm s}},
\end{equation}  
where $M_{\rm s}$ is the stellar mass in a specific cell and $\triangle M_{\rm s}$ the mass of stars
born in the same cell during timestep $\triangle t$. Equation~(\ref{rejuv}) satisfies  the expected
asymptotic behaviour of the stellar age, which becomes small (${\overline\tau}_\ast^{\rm rjv} \rightarrow \triangle t$) during a massive instantaneous burst 
($\triangle M_{\rm s} \rightarrow \infty$), but remains almost unchanged 
(${\overline\tau}_\ast^{\rm rjv} \rightarrow {\overline\tau}_\ast$)  in the quiescent phase
$\triangle M_{\rm s} \rightarrow 0$.  
\item The mean stellar age should uniformly increase with time,
  reflecting the overall aging of the system. This is done by 
  adding the current timestep
$\triangle t$ to the mean stellar age in every grid cell. Thus, in the
absence of star formation and pre-existing old stellar population, the
mean stellar age is equivalent to the current evolution time $t$.
\end{enumerate}

In practice, we first solve for Equation~(\ref{advectTau}) to account for bulk motions, then we update
the mean stellar age in every grid cell according to Equation~(\ref{rejuv})
to account for star formation, and finally we add the current timestep $\triangle t$ to
the mean stellar age in every grid cell to take into account the aging of stellar populations. 
Since we follow the evolution of both the
massive and intermediate-mass star subcomponents, their mean ages
${\overline\tau}^{\rm h}_\ast$ and ${\overline\tau}^{\rm m}_\ast$, 
respectively, need to be calculated using the same procedure.

\begin{figure}
  \resizebox{\hsize}{!}{\includegraphics{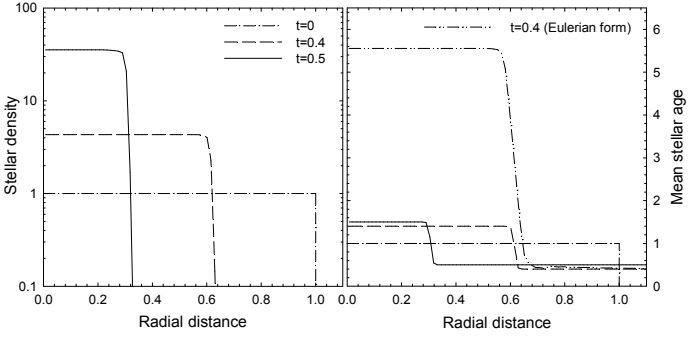}}
  \caption{Star volume density (left) and mean stellar age (right) as a
    function of time during the gravitational collapse of a stellar
    sphere of unit radius. The elapsed time is indicated in the left
    panel. Star formation is turned off for this test problem. The dash-dot-doted
    lines shows the mean stellar age when calculated using the Eulerian form to account for bulk
    motions of stars instead of the Lagrangian form (see text for more detail.) }
         \label{fig1}
\end{figure}

Figure~\ref{fig1} shows the time evolution of the stellar density
$\rho_{\rm s}$ (left panel) and mean stellar age
${\overline\tau}_\ast$ (right panel) in an idealized test problem
involving the gravitational collapse of a stellar sphere with unit
radius.  Initially, $\rho_{\rm s}$ and ${\overline\tau}_\ast$ are set
to unity inside the sphere and to zero outside the sphere (dash-dotted
lines). The stellar velocity dispersion is negligible everywhere and
the star formation rate is set to zero.  As the
collapse proceeds, stellar density shows the expected behaviour with a
central plateau shrinking in size and growing in density. At the same
time the mean stellar age increases in concordance with time, as
expected in the absence of star formation.  In particular, the
difference between the values of ${\overline\tau}_\ast$ inside and
outside the sphere is always equal to unity, as was set initially.
 We emphasize here the need for using the Lagrangian equation
for ${\overline\tau}_\ast$. If the Eulerian form 
$\partial {\overline\tau}_\ast / \partial t+ {\bl\nabla} \cdot 
({\bl v}_{\rm s} {\overline\tau}_\ast)=0 $   
 is used instead, the above test fails. Indeed, the dash-dot-dotted line 
shows the mean stellar age at $t=0.4$ calculated using the Eulerian form.
Evidently, the difference between the values of ${\overline\tau}_\ast$ inside and
outside the sphere is now much greater than unity, indicating a spurious aging
of the stellar populations inside the sphere\footnote{\bf It is still possible to use the Eulerian form
for the product $\rho_{\rm s} {\overline\tau}_\ast $ instead of just ${\overline\tau}_\ast$. 
We leave the exact derivation of the pertinent  equations for a followup study. }.

Finally, we note that the concept of the mean stellar age has its limitations.
For instance, in the case of a constant SFR and steady galaxy configuration, 
${\overline\tau}^{\rm h}_\ast$ 
approaches a constant value after a few tens of Myr, meaning that the supernova rates
and mass return rates (see equations~(\ref{SNII}) and (\ref{sourceSNII})) 
will be exclusively determined by stars whose lifetime is equal to  ${\overline\tau}^{\rm h}_\ast$.
This bias towards stars with single age (and mass) diminishes, when a time-varying
SFR is present or stellar motions are taken into account. Nevertheless, a  more sophisticated
approach that takes into account a possible age spread
around the mean value is desirable.

\subsection{Source and sink terms}
\label{SSterms}
In order to calculate the rates at which stars return their mass
(including newly synthesized elements) into the gas phase, we need to
make assumptions about the initial mass function (IMF), stellar
lifetimes and nucleosynthesis rates.

We adopt the Kroupa IMF \citep{Kroupa2001} of the form
\begin{equation}
\phi(M_\ast)  = \left\{ \begin{array}{ll} 
   A \,  M_\ast^{-1.3} & \,\, 
   \mbox{if $M_\ast \le 0.5~M_\odot$ } \\ 
   B \, M_\ast^{-2.3} & \,\, 
   \mbox{if $M_\ast  > 0.5~M_\odot$ },  \end{array} 
   \right. 
   \label{function}  
 \end{equation}
 where the normalization constants $A$ and $B$ are calculated as
 described in Sections~\ref{SNtwo} and \ref{SNone}.



The stellar lifetimes are taken from \citet{Padovani93}
\begin{equation}
\tau(M_\ast)  = \left\{ \begin{array}{ll} 
   1.2 \, M_\ast^{-1.85} + 0.003~\mbox{Gyr}& \,\, 
   \mbox{if $M_\ast \ge 7.45~M_\odot$ } \\ 
   10^{f(M_\ast)}~\mbox{Gyr} & \,\, 
   \mbox{if $M_\ast  <  7.45~M_\odot$ },  \end{array} 
   \right. 
   \label{starage}
\end{equation}
where 
\begin{equation}
f(M_\ast)= { \left\{ 0.334 - \sqrt{1.79 - 0.2232 \times \left(7.764 - \log(M_\ast) \right) } 
\right\}   \over 0.1116.  }
\end{equation}
In equation~(\ref{starage}), we have adjusted the transitional value
for the stellar mass (7.45~$M_\odot$) in order to obtain a smooth time
derivative $dM_\ast / d\tau$.

\subsubsection{Supernova type II rates}
\label{SNtwo}
We assume that all stars in the $[8.0~M_\odot : M_\ast^{\rm max}]$
mass range end their life cycle as type II supernovae (SNeII), where
$M_\ast^{\rm max}$ is the upper cut-off mass of the IMF set to
$100~M_\odot$ in our model.  The SNII rate, i.e., the number of SNeII
per unit time, is then defined as
\begin{equation}
{\cal R}_{\rm SNII} = \left\{ \begin{array}{ll} 
   \phi[M_\ast({\overline\tau}^{\rm h}_\ast)]  { dM_\ast({\overline\tau}^{\rm h}_\ast)  \over d\tau } \\
   \hskip 1.2cm  \mbox{if} \,\, \overline\tau^{h}_\ast\in [\tau(100~M_\odot):\tau(8.0~M_\odot)], \\ 
   0 \hskip 1.cm  \mbox{if} \,\, \overline\tau^{h}_\ast \ni [\tau(100~M_\odot):\tau(8.0~M_\odot)],  \end{array} 
   \right. 
\label{SNII}   
\end{equation}
where $M_\ast({\overline\tau}^{\rm h}_\ast)$ is the mass of a massive
star that is about to die as SNII calculated by inverting
equation~(\ref{starage}).  This is the SNII rate following an
instantaneous burst of star formation.  We therefore assume here that
each star formation event generates a single stellar population (SSP).
The superposition of SNII rates coming from many SSPs will then
approximate the SNII rate in the presence of a variable SFR. The
derivative $dM_\ast/d\tau$ is calculated by inverting
equation~(\ref{starage}) and then differentiating it with respect to
time.

The IMF in equation~(\ref{SNII}) has to be normalized according to the
total mass of massive stars $M^{\rm h}_{\rm s,tot}$ in each
computational cell
\begin{equation}
\int \limits_{8.0~M_\odot} \limits^{M_\ast({\overline\tau}^{\rm h}_\ast)} 
\phi(M_\ast) M_\ast \, dM_\ast = M^{\rm h}_{\rm s,tot}.
\end{equation}
We note that the upper limit in the integral is 
not fixed but rather depends on the mean stellar age of massive stars
in a given computational cell.  This is done to take into account the
fact that massive stars with lifetimes smaller than
${\overline\tau}^{\rm h}_\ast$, and hence with mass greater than
$M_\ast({\overline\tau}^{\rm h}_\ast)$, must have exploded as SNeII.
Excluding those stars ensures a correct calculation of the stellar
mass spectrum at any time instant\footnote{$M_\ast({\overline\tau}^{\rm
      h}_\ast)$ should depend on $M^{\rm h}_{\rm s,tot}$ because small
    stellar populations, characterized by a small amount of massive
    stars, could have an IMF truncated at relatively low masses
    \citep{ploe14}.  We neglect this effect in this paper.}.
Since $M^{\rm h}_{\rm s}$ and $M_\ast({\overline\tau}^{\rm h}_\ast)$
are time-varying quantities, the normalization constants in the IMF
need to be updated every time step.

  

The mass return rate (per unit volume) by SNII explosions can now be
defined as
\begin{equation}
{\cal S}_{\rm SNII}^{i} = {\cal R}_{\rm SNII} \,\, \rho^i_{\rm SNII}
[M_\ast(\overline\tau^{\rm h}_\ast)],
\label{sourceSNII}
\end{equation}
where $\rho^i_{\rm SNII}[M_\ast(\overline\tau^{\rm h}_\ast)]$ is the
mass (per unit volume) released by a SNII of mass $M_\ast$ in the form
of a specific element $i$.  The values of $\rho^i$ for nine considered
elements (H, He, C, O, N, Ne, Mg, Si, and Fe) and four stellar
metallicities ($Z_\ast=10^{-4}, 10^{-2}, 10^{-1}$, and 1.0  times
solar) are taken from \citet{WW}. For stellar masses greater than
$40~M_\odot$, the yields are set constant to those of a $40~M_\odot$ star.


Finally, we assume that each supernova releases $\epsilon_{\rm SN} \times 10^{51}$~ergs
in the form of thermal energy and define the rate of energy density 
released by SNeII as
\begin{equation}
\Gamma_{\rm SNII}= 10^{51} \epsilon_{\rm SN} {\cal R}_{\rm SNII}\, V^{-1},
\label{SNIIen}
\end{equation}
where $V$ is the injection volume specified by the actual size of our
numerical grid and  $\epsilon_{\rm SN}$ is the ejection efficiency set to 1.0
in this work (see Section~\ref{caveats} for discussion.).

\subsubsection{Supernova type Ia rates}
\label{SNone}
We derive the rates of Type Ia supernovae following the ideas and
notations laid out in \citet{MR01}. The number of binary systems (per
unit mass of the binary system $M_{\rm B}$) that are capable of
producing a SNIa explosion can be calculated as
\begin{equation}
N_{\rm SNIa}[M_2({\overline\tau}^{\rm m}_\ast)]= A_\ast \int \limits_{M_{\rm B,1}} 
\limits^{M_{\rm B,2}} \phi(M_{\rm B}) 
f\left( {M_2({\overline\tau}^{\rm m}_\ast) \over M_{\rm B}}  \right) {dM_{\rm B} \over M_{\rm B}},
\label{SNIa}
\end{equation}
where $M_2({\overline\tau}^{\rm m}_\ast)$ is the mass of the secondary
component, $A_\ast$ a normalization constant and
the limits of the integral are defined as
\begin{eqnarray}
M_{\rm B,1}&=&\max[2M_2({\overline\tau}^{\rm m}_\ast),M_{\rm B,min}] \\
M_{\rm B,2}&=&{1\over 2} M_{\rm B,max} + M_2({\overline\tau}^{\rm m}_\ast).
\end{eqnarray}
The maximum and minimum masses of a binary system that is capable of
producing a SNIa explosion are set to $M_{\rm B,min}=3.0~M_\odot$ and
$M_{\rm B,max}=16.0~M_\odot$, respectively, and the maximum mass of
the secondary is $M_2=8.0~M_\odot$. The quantity
$f(M_{2}({\overline\tau}^{\rm m}_\ast)/M_{\rm B})$ in
equation~(\ref{SNIa}) is the distribution function of the
$M_2({\overline\tau}^{\rm m}_\ast)/M_{\rm B}$ ratio and is defined as
\begin{equation}
f\left( {M_2({\overline\tau}^{\rm m}_\ast) \over M_{\rm B}} \right)=2^{1+\gamma} (1+\gamma) 
\left( {M_2({\overline\tau}^{\rm m}_\ast) \over M_{\rm B}} \right)^\gamma,
\end{equation}
and $\gamma$ is set to 0.5.  For this value of $\gamma$, the
normalization constant that bettter reproduces the observed
[$\alpha$/Fe] ratios in dwarf galaxies is $A_\ast=0.06$
\citep{Recchi09}, therefore we adopt this value in the present work.

By analogy to the massive star subcomponent, the IMF in
equation~(\ref{SNIa}) has to be normalized according to the total mass
of intermediate-mass stars $M^{\rm m}_{\rm s,tot}$ in each
computational cell
\begin{equation}
\int \limits_{1.0~M_\odot} \limits^{M_\ast({\overline\tau}^{\rm m}_\ast)} 
\phi(M_\ast) M_\ast \, dM_\ast = M^{\rm m}_{\rm s,tot}.
\end{equation}

The SNIa rate, i.e., the number of SNIa explosions per unit time, 
can now be calculated by analogy to equation~(\ref{SNII}) as 
\begin{equation}
{\cal R}_{\rm SNIa}  = \left\{ \begin{array}{ll} 
   N_{\rm SNIa}[M_2({\overline\tau}^{\rm m}_\ast))]  { dM_2({\overline\tau}^{\rm m}_\ast)) 
   \over d\tau }  \\
   \hskip 1.2cm \mbox{if} \,\, {\overline\tau}^{\rm m}_\ast\in 
   [\tau(8.0~M_\odot):\tau(M_{\rm 2,min})], \\ 
   0 \hskip 1cm  \mbox{if} \,\, {\overline\tau}^{\rm m}_\ast\ni 
   [\tau(8.0~M_\odot):\tau(M_{\rm 2,min})],  \end{array} 
   \right. 
\label{SNIaRate}
\end{equation}
where $M_{\rm 2,min}=1.0$ M$_\odot$ is the minimum mass of the
secondary. 


The mass return rate (per unit volume) by SNIa explosions can now be
expressed as
\begin{equation}
{\cal S}_{\rm SNIa}^{i} = {\cal R}_{\rm SNIa} \, \rho^i_{\rm SNIa}[M_\ast({\overline\tau}^{\rm m}_\ast)],
\label{sourceSNIa}
\end{equation}
where $\rho^i_{\rm SNIa}[M_\ast({\overline\tau}^{\rm m}_\ast)]$ is the
mass (per unit volume) released by a SNIa of mass $M_\ast$ in the form
of a specific element $i$.  The values of $\rho^i$ for six elements
(C, O, N, Mg, Si, and Fe) are taken from \citet{Nomoto84}. We neglect
the metallicity dependence of the SNIa yields.  The energy release
rate by SNIa ($\Gamma_{\rm SNIa}$) is defined in analogy to
Equation~(\ref{SNIIen}), with ${\cal R}_{\rm SNII}$ being substituted
by ${\cal R}_{\rm SNIa}$.

\subsubsection{ Stellar winds and the mass return by intermediate-mass stars } 
The luminosity produced by stellar winds is taken
from results of the Starburst99 software package \citep{leit99, vl05}.
We used Padova AGB stellar tracks at different metallicities
($Z_\ast$=0.0004, 0.004, 0.008, 0.02, 0.05) and used this set of
models as a library to obtain wind luminosities for each SSP,
depending on its mass and average metallicity.  
The energy transfer efficiency of the wind power is set to 100\%.

%

The mass-return rate (per unit volume) by winds of stars in the $(1.0-8.0)~M_\odot$ mass range 
is calculated as
\begin{equation}
{\cal S}^i_\ast={\cal R}_\ast \rho^i_\ast \left[M_\ast( {\overline\tau}^{\rm m}_\ast)\right],
\end{equation}
where the number of stars with mass $1.0<M_\ast <8.0~M_\odot$ dying per unit
time is expressed as
\begin{equation}
{\cal R}_\ast  = \left\{ \begin{array}{ll} 
   \phi[M_\ast({\overline\tau}^{\rm m}_\ast)]  { dM_\ast({\overline\tau}^{\rm m}_\ast) \over d\tau }  \\
   \hskip 1.2cm  \mbox{if} \,\, t\in [\tau(8.0~M_\odot):\tau(1.0~M_\odot)], \\ 
   0 \hskip 1.cm  \mbox{if} \,\, t\ni [\tau(8.0~M_\odot):\tau(1.0~M_\odot)],  \end{array} 
   \right. 
\label{lms}  
\end{equation}
and the mass (per unit volume) returned by dying stars $\rho^i_\ast$
in the form of a specific element $i$ includes both the true
nucleosynthesis yields of C, O, and N and the pre-existed contribution
\citep[see][]{Recchi01}. The nucleosynthesis yields are taken from
\citet{vdhg97}.

\subsubsection{Star-formation rate}
\label{subs:starformationrate}
The phase transformation of gas into stars is controlled by the
following SFR per unit volume
\begin{equation}
{\cal D}  = \left\{ \begin{array}{ll} 
    c_\ast{(\rho_{\rm g}\xi_{\rm cpw})^{1.5} \over  \mathrm{max}(\tau_{\rm ff},\tau_{\rm c})} 
\left[ T_0^2 \over T_0^2 + T^2\right]^2 \,\, \mathrm{if}~n_{\rm g}\ge n_{\rm cr} 
~\mathrm{and}~{\bl \nabla}\cdot {\bl v}_{\rm g}< 0, \\ 
   0 \hskip 3.2 cm \,\, \mbox{if}~n_{\rm g}< n_{\rm cr}~\mathrm{or}~{\bl \nabla}\cdot {\bl v}_{\rm g} > 0 ,    \end{array} 
   \right. 
\label{sfr}  
\end{equation}
where $n_{\rm g}$ and $T$ are the gas number density and temperature,
respectively, ${\bl \nabla}\cdot {\bl v}_{\rm g}$ is the gas
velocity divergence, and $\xi_{\rm cpw}$ is the cold plus warm gas fraction 
(see Section~\ref{subs:overcooling} for detail).  
The exponent 1.5 in Equation~(\ref{sfr}) is expected
if the SFR is proportional to the ratio of the local gas volume density 
to the free-fall time, and the
free-fall and cooling times are calculated as $\tau_{\rm ff} = \sqrt{3
  \pi /(32 G \rho_{\rm g})}$ and $\tau_{\rm c} = \epsilon/\Lambda$,
respectively\footnote{Strictly speaking, this is only true when $\tau_{\rm c}>\tau_{\rm ff}$. Otherwise, the density exponent should be set to 1.0.}.  The term in brackets applies the adopted temperature
dependence by quickly turning off star formation at temperatures
greater than $T_0=10^3$~K. For consistency with other studies
\citep[e.g.][]{Springel2003}, the critical gas number density above
which star formation is allowed is set to 1.0~cm$^{-3}$ (see discussion
in Section~\ref{caveats}). Finally, the
normalization constant $c_\ast$ is set to 0.05 \citep{Stin06}.

\subsection{Stellar metallicity}
The heavy element yields by SNeII and also by low- and intermediate-mass
stars are metallicity dependent. We calculate the stellar metallicity
$Z_\ast$ using a procedure similar to that applied to the mean stellar age 
in Section~\ref{meanstarage}. First, we solve for the Lagrangian equation for 
$Z_\ast$ to account for bulk motions
\begin{equation}
\label{advectZ}
{d Z_\ast \over d t} \equiv { \partial Z_\ast \over \partial t} +
({\bl v}_{\rm s} \cdot {\bl \nabla }) Z_\ast =0. 
\end{equation}  
Then, we update the stellar metallicity in every grid cell 
to account for the production of metals due to star formation 
using the following equation
\begin{equation}  
\label{newZ}
Z_\ast^{\rm new}={ M_{\rm s} Z_\ast +
\triangle M_{\rm s} \, Z_{\rm g} \over M_{\rm s} +\triangle M_{\rm s}},
\end{equation}  
where $Z_{\rm g}$ is the current metallicity of gas from which stars form.
Equation~(\ref{newZ}) satisfies  the expected
asymptotic behaviour of the stellar metallicity, which approaches that of the gas 
($Z_\ast^{\rm new} \rightarrow Z_{\rm g}$) during a massive instantaneous burst 
($\triangle M_{\rm s} \rightarrow \infty$), but remains almost unchanged 
($Z_\ast^{\rm new} \rightarrow Z_\ast$)  in the quiescent phase
$\triangle M_{\rm s} \rightarrow 0$. We note that step 3 from Section~\ref{meanstarage}
needs not be applied to the stellar metallicity, because this step 
takes into account the overall aging of stellar populations.


\section{Cooling and heating rates}
\label{coolheat}
We assume that the gas is in collisional ionization equilibrium and
calculate the gas cooling rates using the cooling functions presented
in \citet{Boehringer89} and \citet{Dalgarno72}.

\subsection{Low-temperature cooling}
For the low-temperature cooling $T<10^4$ K, we adopted the cooling
rates as described in \citet{Dalgarno72}. The chemical elements that
are taken into account are H, O, C, N, Si, and Fe. For high fractional
ionization, cooling is mostly provided by collisions of neutral and
singly-ionized ions with thermal electrons. We have assumed that most
of oxygen and nitrogen is in neutral form. 
The following equations describe the
cooling efficiency $L_{\rm e}(X_i)$ (in erg~cm$^3$~s$^{-1}$) due
to collisions of element $X_i$ of number density $n_{i}$ with thermal
electrons of number density $n_{\rm e}$.

\begin{eqnarray}
L^{\rm low}_{\rm e}(\mathrm{C}^{+}) &=& T^{-1/2} \left( 7.9\times 10^{-20} \mathrm{e}^{-92/T} \right. \nonumber
\\ &+& \left. 3.0\times 10^{-17} \mathrm{e}^{-61900/T} \right), \\
L^{\rm low}_{\rm e}(\mathrm{Si}^{+}) &=& T^{-1/2} \left( 1.9\times 10^{-18} \mathrm{e}^{-413/T}  \right.
\nonumber \\
&+& \left.  3.0\times 10^{-17} \mathrm{e}^{-63600/T} \right), \\
L^{\rm low}_{\rm e}(\mathrm{Fe}^{+}) &=& T^{-1/2} \left( 4.8\times 10^{-18} \mathrm{e}^{-2694/T} \right.
\nonumber \\ 
&+& 1.1\times 10^{-18} \left[ \mathrm{e}^{-554/T} +1.3\mathrm{e}^{-961/T} \right]  \nonumber \\
&+& \left. 7.8\times10^{-18} \mathrm{e}^{-3496/T} \right), \\
L^{\rm low}_{\rm e}(\mathrm{O}) &=& 1.74\times 10^{-24} T^{1/2} \nonumber \\
&\times& \left[ \left(1-7.6 T^{-1/2}\right) \mathrm{e}^{-228/T} \right. \nonumber \\
&+& \left. 0.38 \left(1-7.7 T^{-1/2}\right) \mathrm{e}^{-326/T} \right] \nonumber \\
&+& 9.4\times 10^{-23} T^{1/2} \mathrm{e}^{-22700/T}, \\
L^{\rm low}_{\rm e}(\mathrm N) &=& 8.2\times 10^{-22} T^{1/2} \nonumber \\
&\times& \left(1-2.7\times10^{-9} T^2\right) \mathrm{e}^{-27700/T}. 
\end{eqnarray}
The cooling rate per unit volume $\Lambda^{\rm low}$ (erg~cm$^{-3}$~s$^{-1}$) due to
collisions of element $X_i$ with thermal electrons is then calculated
as $L^{\rm low}_{\rm e}(X_i) \, n_{\rm i}\, n_{\rm e}$.  Cooling due
to collisions of hydrogen atoms with thermal electrons, as tabulated
in Table~2 of \citet{Dalgarno72} is also taken into account.

For low fractional ionization, collisions of C$^{+}$, O, Si$^{+}$, and
Fe$^{+}$ with neutral hydrogen may become important contributors to
the total cooling budget. The following equations describe the
corresponding cooling efficiency for Si$^{+}$ and Fe$^+$.
\begin{eqnarray}
L^{\rm low}_{\rm H}(\mathrm{Si}^{+}) &=& 7.4\times 10^{-23} \mathrm{e}^{-413/T}, \\
L^{\rm low}_{\rm H}(\mathrm{Fe}^{+}) &=& 1.1\times 10^{-22} \left( \mathrm{e}^{-554/T} +1.4 \mathrm{e}^{-961/T}
\right).
\end{eqnarray}
The cooling rate per unit volume due to collisions of element $X_i$
with hydrogen atoms of number density $n_{\rm H}$ is then calculated
as $L^{\rm low}_{\rm H}(X_i) \, n_i \, n_{\rm H}$.  The cooling
efficiencies due to collisions of O and C$^{+}$ with neutral hydrogen
are taken from Table~4 of \citet{Dalgarno72}.

To summarize, the total cooling rate at the low-temperature regime is
calculated as
\begin{equation}
\label{coollow}
\Lambda^{\rm low}\left[ {\mathrm{erg \over cm^3~s}} \right]=\sum \limits_i n_{i} n_{\rm H} \xi_{\rm
cpw}
\left[ L^{\rm low}_{\rm e}(X_i) \, f_{\rm ion} + L^{\rm low}_{\rm H}(X_i) \, 
 \right],
\end{equation}
where summation is done over six elements: H, O, C, N, Si, and Fe,
where applicable.  Here, $\xi_{\rm cpw}$ is the cold plus warm gas fraction, 
and  $f_{\rm ion}=n_{\rm e}/n_{\rm H}$ is the
ionization fraction, the value of which for a given gas temperature
$T_{\rm g}$ is taken from table~2 of \citet{Schure09} for the
collisionally ionized gas in thermal equilibrium for $T_{\rm
  g}>10^{3.8}$~K. At lower temperatures, $f_{\rm ion}$ is set to a
constant value of 0.01.

We note that most of the considered elements (C, Si, Fe, and  O) 
have relatively low critical densities $n_{\rm crit}$ at which spontaneous
and collisional de-excitations become comparable for some transition
levels. Cooling due to these elements at $n_{\rm H}\ga n_{\rm crit}$ is
proportional to $n_{i} n_{\rm crit}$ rather than to $n_i n_{\rm H}$. We
take this into account by multiplying Equation~(\ref{coollow}) with
$n_{\rm crit}/(n_{\rm crit}+n_{\rm H})$ and taking $n_{\rm crit}$ from
table~8 in \citet{McKee89}. We not that at a low numerical resolution
the critical densities may never be reached.

\subsection{High-temperature cooling}
For the solar-metallicity plasma with temperature $T\ge10^4$~K, we
have adopted the cooling rates of \citet{Boehringer89}, which include
the line emission for most abundant elements and continuum emission
due to free-free, two-photon, and recombination radiation. We have
considered nine chemical elements (H, He, C, N, O, Ne, Mg, Si, and
Fe), the total (line plus continuum) cooling efficiencies $L^{\rm
  high}(X_i)$ for which are plotted in Fig.~2 of \citet{Boehringer89}.
These cooling efficiencies are normalized to $n_{\rm e} n_{\rm H}$ and
hence need to be multiplied by the ionization fraction $f_{\rm
  ion}=n_{\rm e}/n_{\rm H}$ to retrieve the values that can later be
used in our hydro code.  The total cooling rates in the
high-temperature regime is then calculated as
\begin{equation}
\Lambda^{\rm high} \left[ \mathrm{erg \over cm^3~s} \right] = \sum \limits_i L^{\rm high}(X_i) \, f_{\rm
ion} \, (n_{\rm H}\xi_{\rm cpw})^2 {n_{i} \,\over n_i^{\rm sol}},
\label{coolhigh}
\end{equation}
 where $n_i$ is the current number density of element $i$ , $n_i^{\rm sol}$ is 
its number density at the solar abundance,
and summation is performed over nine elements: H, He, C,
N, O, Ne, Mg, Si, and Fe.

\begin{figure}
  \resizebox{\hsize}{!}{\includegraphics{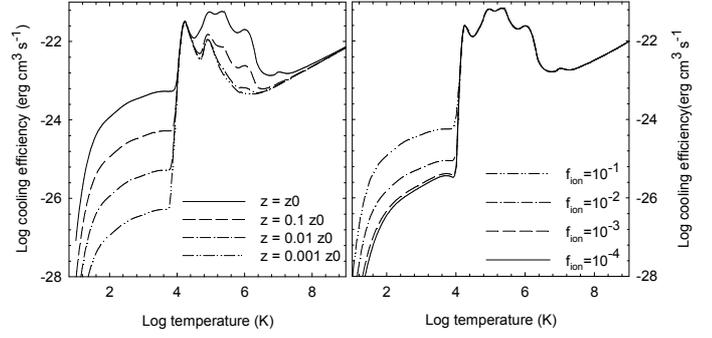}}
  \caption{Cooling efficiency of gas as a function of temperature for
    the ionization fraction $f_{\rm ion}=1.0$ but varying metallicity
    (left panel) and for the gas with solar metallicity but varying
    ionization fractions $f_{\rm ion}$ (right column).}
         \label{fig2}
\end{figure}

Figure~\ref{fig2} shows the cooling efficiency laid out in the following form
\begin{equation}
\sum \limits_i {n_i\over n_H} 
\left[ f_{\rm ion} L^{\rm low}_{\rm e}(X_i)  +L^{\rm low}_{\rm H}(X_i) \right] 
+f_{\rm ion} L^{\rm high}(X_i).
\end{equation}
In particular, the left panel has $f_{\rm ion}=1$ but varying
metallicity and the right panel has the solar metallicity but varying
$f_{\rm ion}$.  A strong dependence on the metallicity and ionization
fraction for $f_{\rm ion}\ga 10^{-3}$ is evident in the figure.

\subsection{Gas temperature}
In order to use the cooling rates described above, one needs to know
the gas temperature $T$.  The latter is calculated form the ideal
equation of state $P=\rho_{\rm g} {\cal R} T_{\rm g}/ \mu$, where
${\cal R}$ is the universal gas constant, using the following
expression for the mean molecular weight
\begin{equation}
\mu={\sum \limits_i \rho_{i}  + m_{\rm e} f_{\rm ion} n_{\rm H} \over \sum \limits_i n_i +  f_{\rm ion} n_{\rm H} },
\end{equation}
where $m_{\rm e}$ is the electron mass and the summation is performed
over all nine chemical elements considered in this study.

\subsection{Avoiding the overcooling of SN ejecta}
\label{subs:overcooling}
When the supernova energy is released in the form of the thermal
energy, there may exists the risk of overestimating the gas cooling,
which may lead to radiating away most of the released thermal energy. This occurs
due to the fact that the cooling rate $\Lambda$ is proportional to the
square of gas number density, which, for a single-phase medium, may be
significantly higher than what is expected in the case of a
multi-phase medium for the hot supernova ejecta.

To avoid the spurious gas overcooling, we make use of the following
procedure. We keep track of the hot SN ejecta by solving for the
following equation for the hot gas density $\rho_{\rm g,h}$
\begin{equation}
{\partial \rho_{\rm g,h} \over \partial t} + {\bl \nabla} \cdot \left( \rho_{\rm g,h}
{\bl v}_{\rm g} \right) = {\cal S} - {\rho_{\rm g,h} \over \tau_{\rm c,h}},
\label{hotgas}
\end{equation}
The second term on the r.h.s. accounts for the phase transformation of
the hot gas into the cold plus warm (CpW) component, where $\tau_{\rm
  c,h}=\epsilon/\Lambda_{\rm h}$ denotes the cooling time of the hot
component and the cooling rate $\Lambda_{\rm h}$ is calculated
assuming the characteristic cooling efficiency of $5\times
10^{-23}$~ergs~cm$^3$~s$^{-1}$ of the hot component with the typical
temperature of $10^8$~K 
(see Fig.~\ref{fig2} and Section~\ref{caveats})\footnote{ In a more accurate approach,
$\Lambda_{\rm h}$ can be directly calculated using equation~(\ref{coolhigh})}.

Equation~(\ref{hotgas}) is solved along with the system of
equations~(\ref{contingas})--(\ref{energy}) describing the evolution
of the gas disk and the CpW gas fraction is calculated as $\xi_{\rm
  cpw}=(\rho_{\rm g} - \rho_{\rm g,h})/\rho_{\rm g}$.  We then modify
the cooling rates and the SFR by multiplying the total gas density
$\rho_{\rm g}$ and the hydrogen number density $n_{\rm H}$ in
Equations~(\ref{coollow}) and (\ref{coolhigh}) and also in
Equation~(\ref{sfr}) with $\xi_{\rm cpw}$ to retrieve the density of
the CpW component.  This procedure scales down the cooling rates and
the SF rate according to the mass of the CpW component in each
computational cell and helps to avoid overcooling and unphysically
high SF rates.

\subsection{Heating processes}
\subsubsection{Cosmic ray heating}
The cosmic ray heating efficiency is the product of the cosmic-ray
ionization rate $\xi_{\rm cr}$ and the energy input per ionization
$\triangle Q_{\rm cr}$.  We adopt $\xi_{\rm cr}=4\times
10^{-17}$~s$^{-1}$ based on the work of \citet{Dishoeck86}. The value
of $\triangle Q_{\rm cr}$ is approximately 20 eV. The resulting
heating rate due to ionization of hydrogen and helium atoms is
\begin{equation}
\Gamma_{\rm cr}=1.34\times 10^{-27} G_0 \left( 0.48 n_{\rm H} + 0.5 n_{\rm He} \right) 
\left[{\mathrm{ergs} \over \mathrm{cm}^{3}~\mathrm{s}}\right],
\end{equation}
where the coefficients 0.48 and 0.5 are taken from table~6 in
\citet{McKee89}.  The quantity $G_0$ represents the interstellar
  radiation field (IRF) normalized to the solar neighbourhood value,
i.e. we make here the usual assumption that the cosmic ray
  intensity scales with the IRF.  We calculate its value as
$G_0=\mathrm{SFR}_{\rm loc}/\mathrm{SFR}_{\rm MW}$, where SFR$_{\rm
  loc}$ is the local SFR in our models and SFR$_{\rm
  MW}=1.0~M_\odot$~yr$^{-1}$ is the SFR in the Milky Way
\citep{Robitaille10}.

\subsubsection{Photoelectric heating from small dust grains}
The photoelectric emission from small grains induced by FUV photons
can substantially contribute to the thermal balance of the
interstellar medium at temperatures $\la 10^4$~K. We use the results
of \citet{BT94} and express the photoelectric heating rate per unit
volume as
\begin{equation}
\Gamma_{\rm ph}= 1.0\times 10^{-24} \epsilon_{\rm ph} \, G_0 n_{\rm H} \xi_{\rm d} \,\, \left[ \mathrm{ergs~cm}^{-3} \mathrm{s}^{-1} \right].
\end{equation}
The photoelectric heating efficiency $\epsilon_{\rm ph}$ is calculated
from Figure~13 of Bakes \& Tielens.
We have also introduced the coefficient $\xi_{\rm d}=(M_{\rm d}/M_{\rm
  HI})/(M_{\rm d}/M_{\rm HI})_\odot$ to take into account a (possibly)
different dust to gas mass ratio $(M_{\rm d}/M_{\rm HI})$ in DGs. 
The latter is found using the following relation between the
oxygen abundance (by number density) and the dust-to-gas mass ratio in
dwarf galaxies \citep{Lisenfeld}
\begin{equation}
12 + \log\left({\mathrm{O} \over \mathrm{H}}\right) = 
0.52 \, C_{\rm d} \, \log\left({M_{\rm d} \over M_{\rm HI}} \right),
\end{equation}
where the normalization constant $C_{\rm d}=-5.0$ is derived from the
corresponding quantities of the Small Magellanic Cloud. For the solar
dust to gas mass ratio we adopt $(M_{\rm d}/M_{\rm HI})_\odot=0.02$,
typical for the interstellar medium in the Solar neighbourhood.


\section{Solution method}
\label{solution}
The equations of gas and stellar hydrodynamics are solved in
cylindrical coordinates with the assumption of axial symmetry using a
time-explicit (except for the gas internal energy update due to gas
cooling/heating), operator-split approach similar in methodology to
the ZEUS code \citep{SN92}.  We use the same finite-difference scheme
but, contrary to the ZEUS code, we advect the internal energy density
$\epsilon$ rather than the specific one $\epsilon/\rho_{\rm g}$
because the latter may give rise to undesired oscillations in some
test problems \citep[e.g.][]{Clarke10}.  Below, we briefly review the
numerical scheme. The test results of our author's implementation of
the code are provided in the Appendix.

The solution procedure consists of three main steps. Step 1: we
update the gas and stellar components due to star formation and
feedback.  The corresponding ordinary differential equations
describing the time evolution of $\rho_{\rm g}$, $\rho_{\rm i}$,
$\rho_{\rm g} {\bl v}_{\rm g}$, $\epsilon$, $\rho_{\rm s}$, $\rho_{\rm
  s} {\bl v}_{\rm s}$, and $\Pi_{ij}$ due to the source and sink terms
involving $\cal D$, ${\cal D}^i$, $\cal S$, ${\cal S}^i$, and
$\Gamma_\ast$ in equations~(\ref{contingas})--(\ref{energy}) and
(\ref{continstar})--(\ref{dispstar}) are solved using a first-order
explicit finite-difference scheme.  In Step 1, we also calculate
the total gravitational potential of the gas and stellar components by
solving for the Poisson equation~(\ref{poisson}) using the alternative
direction implicit method described in \citet{BB75}.  The
gravitational potential at the outer boundaries is calculated using a
multipole expansion formula in spherical coordinates
\citep{Jackson75}. To save on computational time, we invoke the
gravitational potential solver only when the relative change in the
total (gas plus stellar) density exceeds 10$^{-5}$ \citep[see][for
details]{SN92}.

Step 2: we solve for the gas hydrodynamics
equations~(\ref{contingas})--(\ref{energy}) excluding the source and
sink terms that have been taken into account in the previous step.
The solution is split into the transport and source substeps. During
the former, advection is calculated using a third-order accurate
piecewise-parabolic scheme of \citet{CW84} modified for cylindrical
coordinates as described in \citet{Blondin93}.  During the latter, we
update the gas momenta $\rho_{\rm g} {\bl v}_{\rm g}$ due to
gravitational and pressure forces and the gas internal energy
$\epsilon$ due to compressional heating using a solution procedure
described in \citet{SN92}.

At the end of Step 2, a fully implicit backward Euler scheme combined
with Newton-Raphson iterations is used to advance $\epsilon$ in time
due to volume cooling  $\Lambda$ and heating $\Gamma$ rates.
In order to monitor accuracy, the total change in $\epsilon$ in one
time step is kept below 10\%. If this condition is not met, we employ
local subcycling in a particular cell by reducing the time step in the
backward Euler scheme by a factor of two (as compared to the global
hydrodynamics time step) and making two individual substeps in the
cell where accuracy was violated. The local time step may be further
divided by a factor of two until the desired accuracy is reached.  The
local subcycling help to greatly accelerate the numerical simulations
as one need not to repeat the solution procedure for every grid cell.

In Step 2, we also add tensor artificial viscosity to
equations~(\ref{momgas}) and (\ref{energy}) to smooth out strong
shocks as described in \citet{SN92}.  As pointed out in \citet{Tasker}
and discussed in \citet{Clarke10}, the ZEUS code with the
\citet{Neumann50} definition of the artificial viscosity performs poorly on
multidimesional tests such as the Sedov point explosion, yielding
non-spherical solutions that may overshoot or undershoot the analytic
one.  We demonstrate in Section~\ref{sedov} that using the tensor
artificial viscosity and the proper choice of the artificial viscosity
parameter $C_2=6$ can greatly improve the code performance. However,
contrary to the Stone \& Norman suggestion to discard the off-diagonal
terms of the viscous stress tensor, we demonstrate that in fact these
terms should be retained and the most general formulation of the
artificial viscosity stress tensor should be used
\begin{equation}
\mathrm{\bf Q} = \left\{ \begin{array}{ll} 
   l^2 \rho_{\rm g} {\bl \nabla} \cdot {\bl v}_{\rm g} 
    \left[ {\bl \nabla} {\bl v}_{\rm g} - \mathrm{\bf e} ({\bl \nabla} \cdot {\bl v}_{\rm g} )/3    \right] &
   \,\,\, \mbox{if ${\bl \nabla} \cdot {\bl v}_{\rm g} < 0$ } \\ 
   0 & \,\,\, 
   \mbox{if ${\bl\nabla} \cdot {\bl v}_{\rm g}\ge0$ },  \end{array} 
   \right. 
   \label{artviscQ}
\end{equation}
where {\bf e} is the unit tensor, $l=C_2 \max(dx)$, and $dx$ is the
grid resolution in each coordinate direction. For the reasons of the
code stability, all components of the viscous sterss tensor {\bf Q}
should be defined at the same position on the grid, i.e., at the zone
centers (contrary to the ZEUS code which defines the off diagonal
components of rank-two tensors at the zone corners).  The Courant
limitations on artificial viscosity may be substantial in regions with
strong shocks. Therefore, we apply local subcycling (in the same
manner as it is done with the internal energy update due to
heating/cooling) when the local viscous time step becomes a small
fraction of the global hydrodynamics time step.

In step 3, we solve for the stellar hydrodynamics
equations~(\ref{continstar})--(\ref{dispstar}) excluding the source
and sink terms that have been taken into account during step 1.  The
solution procedure is similar to that for the gas hydrodynamics
equations, since both systems are defined essentially on the same grid
stencil---a powerful advantage of the stellar hydrodynamics approach.
All components of the velocity dispersion tensor are defined at the
cell centers to be consistent with the definition of the stellar
artificial viscosity tensor {\bf Q}. The latter is defined by analogy
to its gas counterpart with the gas density and velocity in
equation~(\ref{artviscQ}) substituted with the stellar density and
velocity.  Additional terms $-{\bl\nabla} \cdot \mathrm{\bf Q}$ and
$-Q_{ik}:({\bl\nabla} {\bl v}_{\rm s})_{kj} - Q_{jk}:({\bl\nabla} {\bl
  v}_{\rm s})_{ki}$ have been added to the r.h.s. of
equations~(\ref{momstar}) and (\ref{dispstar}), respectively, to take
into account the viscous stress and heating due to artificial
viscosity.


Finally, the global time step for the next loop of simulations is
calculated using the Courant condition modified to take into account
star formation and feedback.  In particular, two time step delimiters
of the form
\begin{eqnarray}
\triangle t_{\rm s.f.}&<&{C_3 \rho_{\rm g}\over {\cal D}}, \,\,\,\mathrm{(star~formation)}, \\ 
\triangle t_{\rm f.b.}&<&{C_3 \rho_{\rm s}\over{\cal S}}, \,\,\,\mathrm{(stellar~feedback)} 
\end{eqnarray}
are added to the usual Courant condition to avoid obtaining negative
values of the flow variables. The safety coefficient $C_3$ is set to
0.5. Some modification to the classic Courant condition is required
for the stellar hydrodynamics part because the stellar velocity
dispersion tensor $\bl \Pi$ is anisotropic in general. To calculate
the stellar hydrodynamics time step, we use the following form
\begin{equation}
\triangle t_{\rm st} = \min \left[ { dx \over |{\bl v}_{\rm s}| +\max \overline{\Pi}_{ii} }  \right],
\end{equation}
where $\max\overline{\Pi}_{ii}$ is the maximum diagonal element of the
diagonalized velocity dispersion tensor and minimum is taken over all
grid zones.

\begin{table*}
\center
\caption{Model parameters}
\label{table1}
\vspace{3 pt}
\begin{tabular}{ccccccc}
\hline\hline
Model & $\alpha$ &  $M_{\rm g}(\le 1~\mathrm{kpc})$ & $M_{\rm g}(\le 3~\mathrm{kpc})$ & 
$M_{\rm g}(\mathrm{total})$ &  self-consistent   & stellar motion    \\
& & $M_\odot$ & $M_\odot$ & $M_\odot$  & initial model &  \\
\hline
1 & 0.25 & $2.1\times 10^7$ & $2.9\times 10^7$ & $3.5\times 10^7$ & yes & yes   \\
2 & 0.5 & $2.8\times 10^7$ & $4.5\times 10^7$ & $6.4\times 10^7$ & yes & yes   \\
3 & 0.9 & $1.1\times 10^8$ & $4.1\times 10^8$ & $1.0\times 10^9$ & yes & yes   \\
4 & 0.5 & $2.8\times 10^7$ & $4.5\times 10^7$ & $6.4\times 10^7$ & yes & no   \\
5 & 0.5 & $1.0\times 10^7$ & $4.5\times 10^7$ & $1.2\times 10^8$ & no & yes   \\
\hline
\end{tabular}
\end{table*}

\section{The chemo-dynamical evolution of model dwarf galaxies}
\label{sec:results}
Our numerical code  has been extensively tested using a standard set
of test problems as described in the Appendix. However, even a perfect
performance on the standardized tests does not guarantee meaningful results 
on real problems when all parts are invoked
altogether. 
 
Therefore, in this section, we consider the evolution of several representative
models of DGs in order to test the overall code performance and to 
demonstrate the simulation methodology and
recipes described in the previous sections.  In particular, we wish to
emphasize the effect of stellar dynamics and initial conditions on the
simulation outcomes.  Moreover, we will show the results of
simulations with different degrees of rotational support against
gravity parameterized by the $\alpha$-parameter defined as
$\alpha=v_\phi/v_{\rm circ}$, where $v_{\rm \phi}$ and $v_{\rm circ}$
are the rotation and circular velocities, the latter being the maximum velocity 
calculated from the assumption that all support against gravity comes from
rotation (i.e., zero gas pressure gradients).  We consider three
different degrees of rotational support: $\alpha$ =0.25, 0.5, 0.9.
The resulting initial geometries vary from flat-like for large values
of $\alpha$ to roundish for small $\alpha$.


\begin{figure}
  \resizebox{\hsize}{!}{\includegraphics{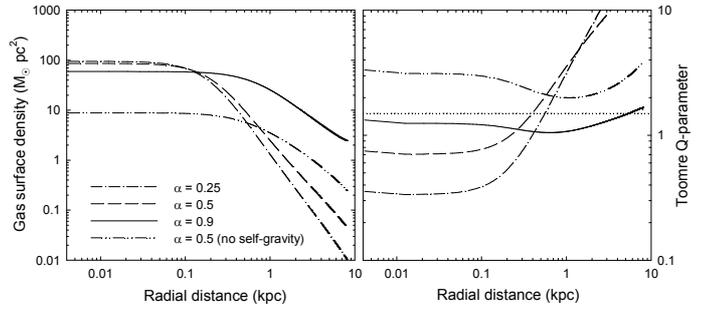}}
  \caption{Initial radial profiles of the gas surface density
    $\Sigma_{\rm g}$ (right) and Toomre $Q$-parameters (left) in
    models with different rotational support against gravity as
    indicated by the $\alpha$-parameter.  The horizontal dotted lines
    define the critical values of $Q$-parameter below which
    star formation is supposed to occur (see the text for more
    details).}
         \label{fig3}
\end{figure}

Our initial configuration consists of a self-gravitating gaseous disk
submerged into a fixed DM halo described by a cored isothermal sphere
\citep[e.g.][]{Silich2001}.  We focus on models with a DM halo mass of
10$^9$ M$_\odot$ and initial gas temperatures of 10$^4$~K. We apply
the solution procedure described in \citet{VRH1} to construct
self-gravitating equilibrium configurations for different values of
$\alpha$.  The initial gas metallicity $Z_{\rm g}$ is set to $10^{-2}
Z_\odot$, the ratio of specific heats
$\gamma$ to 5/3,  and the redshift is set to zero. The number of grid zones 
is 1000 in each coordinate direction, resulting in the effective numerical
  resolution of 8~pc everywhere throughout the computational domain.
  We impose the equatorial symmetry at the midplane and the reflection
  boundary condition at the axis of symmetry. At the other boundaries,
  a free outflow condition is applied.

Figure~\ref{fig3} presents the initial gas surface density
$\Sigma_{\rm g}$ (left panel) and the Toomre $Q$-parameter (right
panel) as a function of the midplane distance in the $\alpha=0.25$
model~1 (dash-dotted lines), $\alpha=0.5$ model~2 (dashed lines), and
$\alpha=0.9$ model~3 (solid lines). Evidently, the $\alpha=0.25$ model
is characterized by the most centrally condensed gas distribution. The
horizontal dotted line marks a  critical $Q$-parameter of 1.5
below which star formation is supposed to occur according to
the stability
analysis of self-gravitating disks by \citet{Toomre1964} and
\citet{Polyachenko1997} \footnote{For a more detailed analysis of
  these star formation criteria see \citet{VRH1}.}.  Evidently, the
$\alpha=0.25$ and $\alpha=0.5$ models are strongly inclined to star
formation in the inner 0.5~kpc, while outside this radius the
conditions are not favourable due to a rather steep drop in the gas
surface density with distance. On the other hand, the $\alpha=0.9$
model is characterized by a much shallower profile of $\Sigma_{\rm
  g}$, resulting in a significantly larger galactic volume prone to
star formation.  The parameters of our models are listed in
Table~\ref{table1}, where $M_{\rm g}(r\le1~\mathrm{kpc})$, $M_{\rm
  g}(r\le 3~\mathrm{kpc})$, and $M_{\rm g}(\mathrm{total})$ are the
gas masses inside a spherical radius of 1~kpc, 3~kpc and the total gas
mass in the computational domain of $8.0\times8.0$~kpc$^2$.

Evidently, the gas mass increases with increasing $\alpha$,
which is explained by the fact that models with higher rotation can support
a higher gas mass against gravity of the DM halo. We note that 
the maximum gas mass is also limited by the action of gas self-gravity \citep{VRH1}.
As the $\alpha=0.5$ model demonstrates, switching off the gas self-gravity 
leads to a factor of two increase in the  total gas mass. 

When constructing the
initial gas density distribution in models~1, 2, and 3, we have chosen the same
``seed'' value for gas number density in the galactic center equal to 10~cm$^{-3}$.
This was done in order to make the initial volume densities and temperatures 
in the galactic center similar in each model. We could have decreased the seed value in models~2
and 3, trying to adjust the total mass in these models to that obtained in model~1, 
but then the initial conditions
for star formation would also have changed. In addition, it was not clear which mass to
take as the reference value, because different values are obtained when integrating over 
different radii due to variations in the radial density profiles.

%

\begin{figure*}
  \hspace{2cm}\resizebox{13cm}{!}{\includegraphics{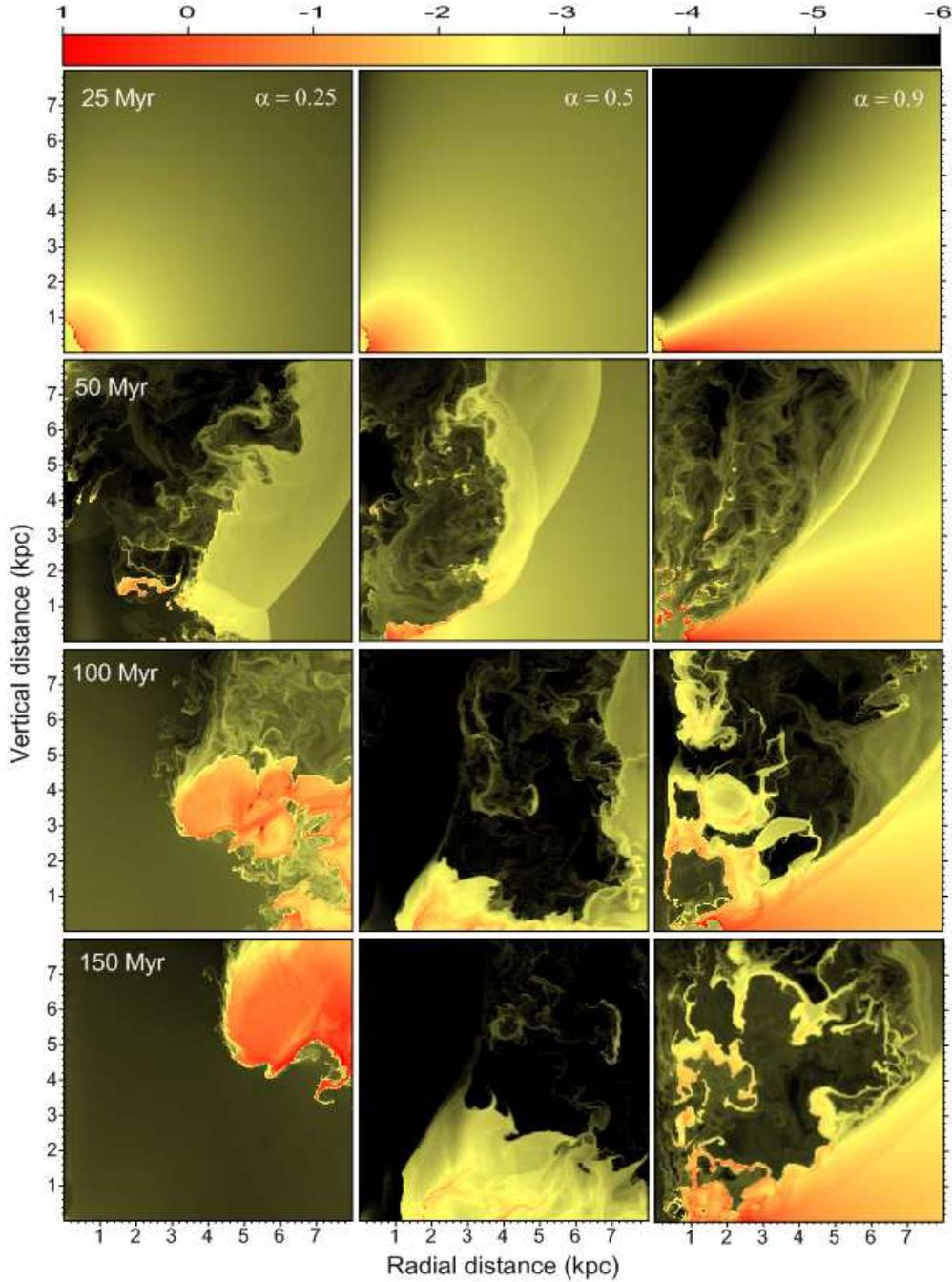}}
  \caption{Gas density distribution of three model galaxies differing
    for the amount of rotational support: the $\alpha$=0.25 model~1
    (left column), the $\alpha$=0.5 model~2 (middle column), and the
    $\alpha$=0.9 model~3 (right column).  The time passed from the
    beginning of numerical simulations is indicated in the leftmost
    image of each row.  The density scale (in cm$^{-3}$) is shown in
    the upper strip.}
         \label{fig3a}
\end{figure*}

\subsection{The role of $\alpha$}
\label{subs:alpha}

In Figure \ref{fig3a} we show a comparison of the gas volume density
($\rho_{\rm g}$) distribution in three model galaxies differing for
the value of $\alpha$.  Four representative time snapshots are taken
for each model.  The $\alpha$=0.9 model~3 is characterized by the largest
degree of flattening, due to the large centrifugal forces.  A notable
flaring in the initial gas density distribution of model~3 can be seen
in the upper-right panel.  This means that the vertical distribution
at $R=0$ is characterized by a very steep density and pressure
gradient.  The over-pressurized superbubble created by the stellar
feedback preferentially expands along this direction and a bipolar
outflow is soon created.  The external parts of the disk are not
affected by the bipolar outflow and even after $t=150$~Myr the gas
density of the disk at radii larger than 1~kpc is quite large.  Some
of the material in the bipolar outflows falls back onto the disk.  As
a result, the fraction of pristine gas lost by this model is
relatively small, in spite of the fact that the galactic outflow is
very prominent and affects a large fraction of the computational
domain.  However, the freshly produced metals can be easily channelled
out of the galaxy.

At the other extreme is the $\alpha$=0.25 model~1.  Here, the initial
gas density configuration is almost spherical and the pressure
gradient is only slightly steeper in the vertical direction than
in the horizontal one.  A bipolar outflow soon develops but
there is significant gas transport also along the disk and, after
100~Myr, all the gas in the central part of the galaxy (up to a radius
of almost 5~kpc) has been completely blown away.

\begin{figure}
  \resizebox{\hsize}{!}{\includegraphics{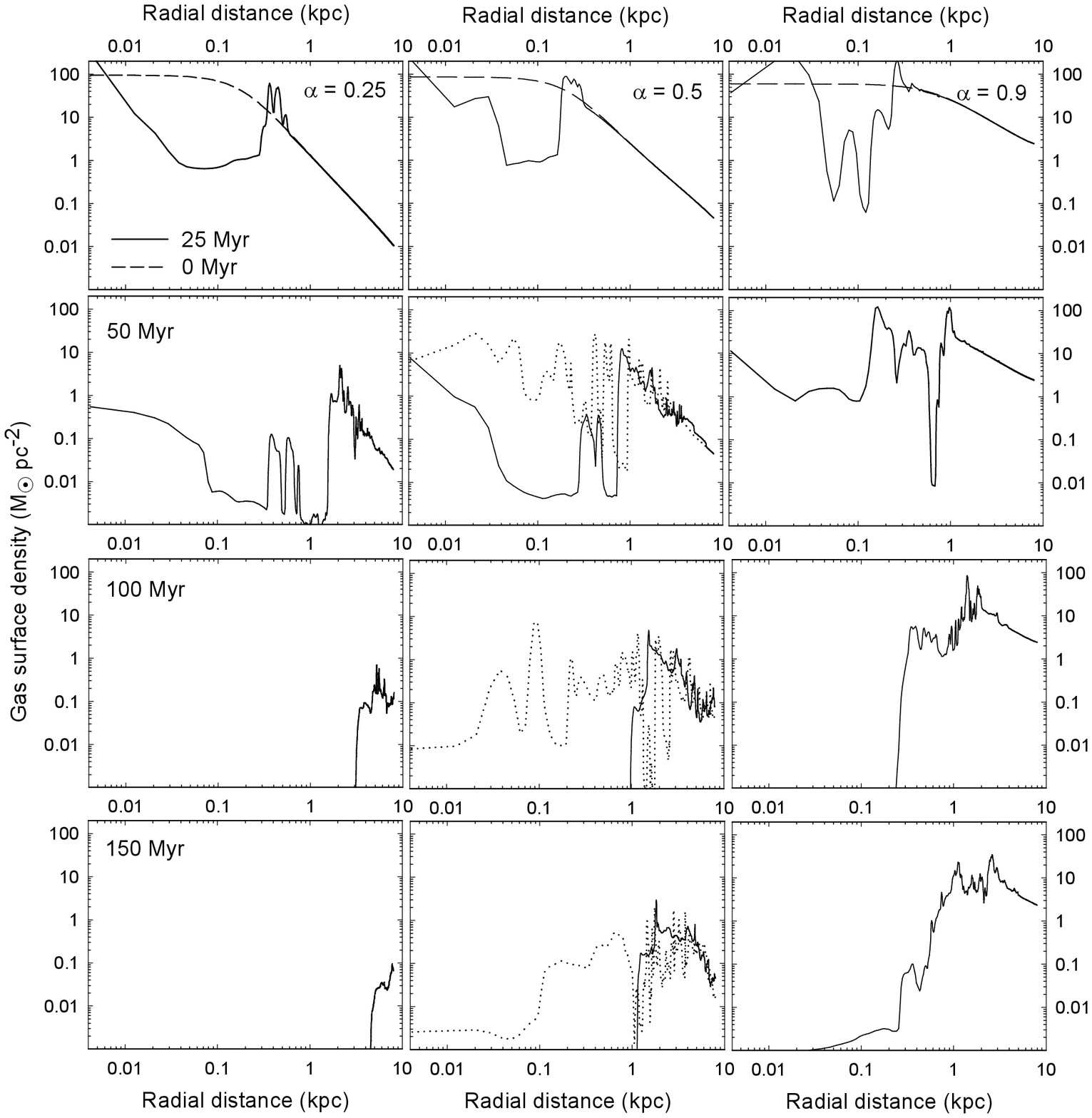}}
  \caption{Gas surface density as a function of radius for the same
    models and at the same moments in time as in Figure \ref{fig3a}.
    In the upper panels the initial density distribution is also
    indicated (dashed lines). The dotted lines in the middle column
    also show the distribution for the $\alpha=0.5$ model without
    stellar motion (see Section~\ref{subs:parameters}). }
         \label{fig4}
\end{figure}

This effect is illustrates in Figure~\ref{fig4}, where the gas surface
density as a function of radius is plotted for our three reference
models at the same moments in time as in Figure~\ref{fig3a}.  After
100~Myr there is almost no gas left in the inner 3~kpc in the
$\alpha$=0.25 model~1, whereas a central hole in the
$\alpha$=0.9~model~3 is of a much smaller radius, hardly exceeding
0.2~kpc, and the gas distribution at $R>2$~kpc is nearly undisturbed.
This different galactic evolution, depending on the initial
distribution of gas, has been already described in the literature
\citep[see in particular][] {rh13}, but here we show it in a more
realistic context of self-gravitating initial gas configurations.

\begin{figure}
  \resizebox{\hsize}{!}{\includegraphics{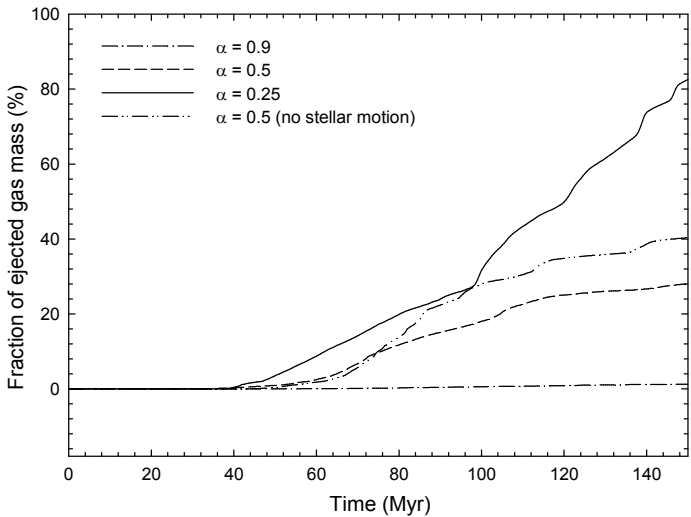}}
  \caption{Fraction of ejected gas mass as a function of time
    for the models with different values of $\alpha$.  The
    dashed-double dotted line refers to a model in which the stellar
    motion has been artificially suppressed (see Sect.
    \ref{subs:parameters}). }
         \label{fig5}
\end{figure}

\begin{figure}
  \resizebox{\hsize}{!}{\includegraphics{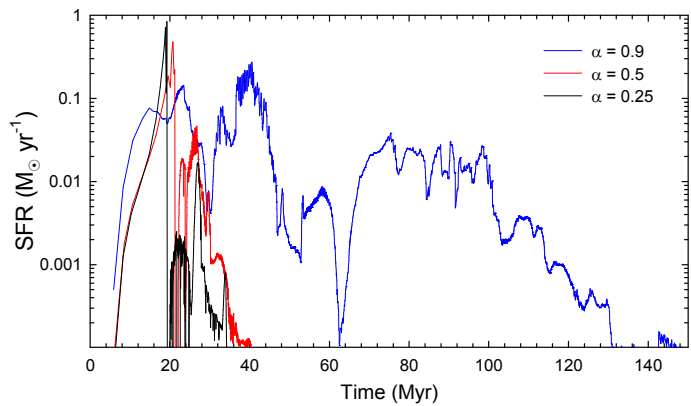}}
  \caption{Star formation rates vs. time in three models differing in
    the value of the centrifugal support $\alpha$. }
         \label{fig6}
\end{figure}

A direct way to appreciate the role of $\alpha$ on the evolution of
our model galaxies is by calculating the ejected gas mass for various
models.  Figure~\ref{fig5} presents the fraction of ejected gas mass
vs. time in models~1--3.  We find this value by calculating the amount
of gas that leaves the computational domain ($8\times8$~kpc$^2$) with
velocities greater than the escape velocity at the computational
boundary.  The fraction of ejected mass is very large in the
$\alpha$=0.25 model, because of the blow-away described above, but it
is very small for the $\alpha$=0.9 model because only a small fraction
of the gas originally present in the disk is involved in the galactic
outflow. The dash-double dotted line presents the fraction of ejected
mass in the $\alpha=0.5$~model~2, in which the motion of star was
artificially suppressed (see Sect.~\ref{subs:parameters} for details.)

The different gas dynamics of the three reference models shown in
Figure~\ref{fig3a} has dramatic consequences also for the star
formation histories. Figure~\ref{fig6} presents the star formation
rate vs. time.  Evidently, the strong star formation feedback in the
$\alpha$=0.25 model shuts off star formation after $\sim$ 30~Myr. The
$\alpha$=0.5 model behaves similarly, although star formation is shut
off slightly later.  On the other hand, the $\alpha$=0.9 model is able
to sustain star formation for a longer time, mainly because the disk
still remains gas-rich and the conditions for the star formation to
occur are fulfilled in many grid cells.

\begin{figure*}
  \hspace{2cm}\resizebox{13cm}{!}{\includegraphics{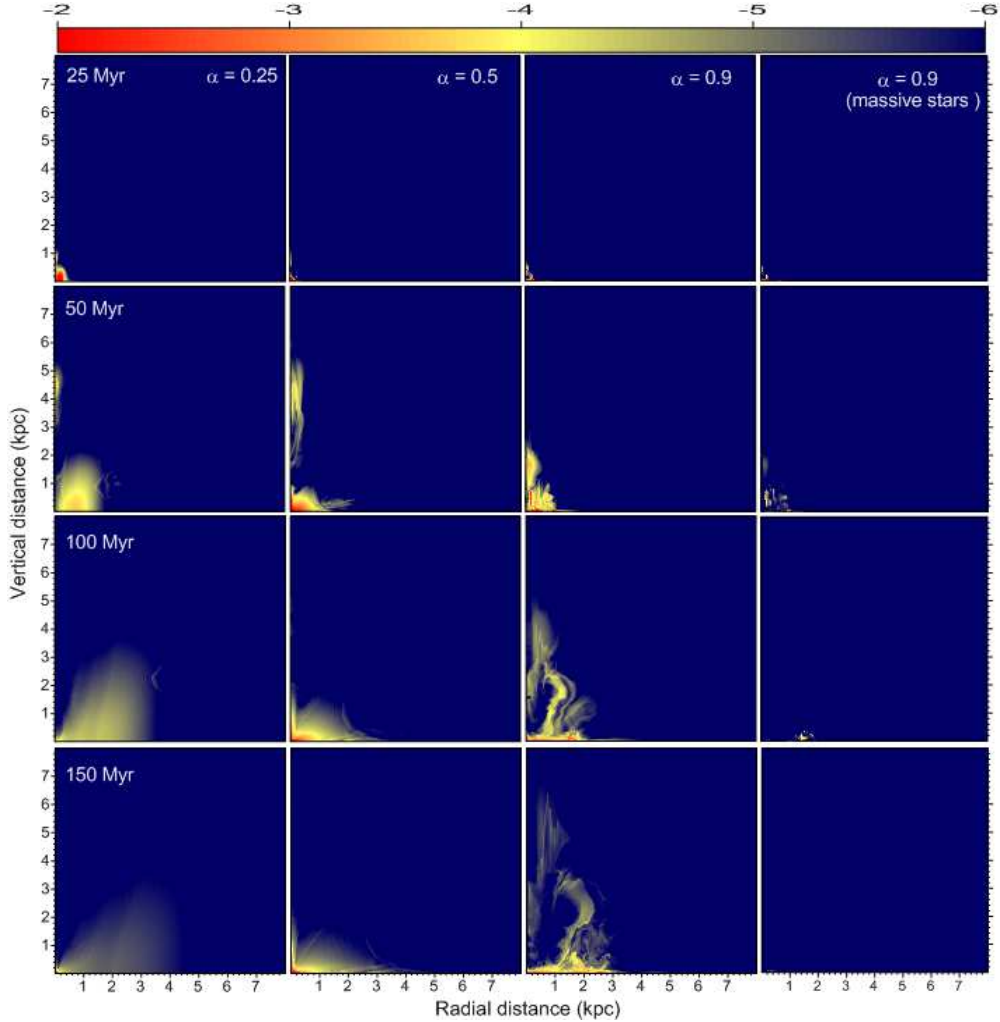}}
  \caption{Similar to Fig. \ref{fig3a}, but now for the stellar volume
    density distribution. The scale bar is in $M_\odot$~pc$^{-3}$.
The right-hand column presents only the massive stars.}
         \label{fig7}
\end{figure*}

Figure~\ref{fig7} presents the distribution of the stellar volume
density $\rho_{\rm s}$ as a function of time and for the same models
and at the same time instances as in Figure~\ref{fig3a}. In addition, 
the right-hand column presents the volume density of high-mass stars
$\rho_{\rm s}^{\rm h}$. We can
clearly see that the stellar distribution in the $\alpha=0.25$ model
is more homogeneous and isotropic than in the $\alpha=0.9$ model,
wherein many stars tend to {\rm be located} along the polar axis.  We note,
however, that owing to a preferable concentration of gas toward the
midplane in the $\alpha$=0.9 model, a significant fraction of stars
is also distributed  near the midplane, constituting a stellar
disk. This effect is much less evident for the models with
$\alpha$=0.5 and $\alpha$=0.25. To quantify this, we calculated the
fraction of the total stellar mass confined within a vertical height
of 1~kpc and 2~kpc.  It turns out that in the $\alpha=0.25$ model
these values are 37\% and 62\%, respectively, whereas in the
$\alpha=0.9$ model they are 90\% and 94\%, indicating a much more
compact stellar distribution in the higher-$\alpha$ model.
 Finally, a comparison of the distribution between $\rho_{\rm s}$ (the total
stellar density) and $\rho_{\rm s}^{\rm h}$ in the $\alpha$=0.9 model indicates that
stars are preferentially forming at low galactic altitudes $\la 1$~kpc, but are later spread out 
owing to their own initial dispersion (5~km~s$^{-1}$) and, especially, owing to 
significant bulk velocities inherited from parental gas clouds.
The off-plane star formation is taking place predominantly in dense gaseous 
clumps and on the cavity walls.

\subsection{The role of the stellar motion and the initial gas distribution}
\label{subs:parameters}

\begin{figure*}
  \hspace{2cm}\resizebox{13cm}{!}{\includegraphics{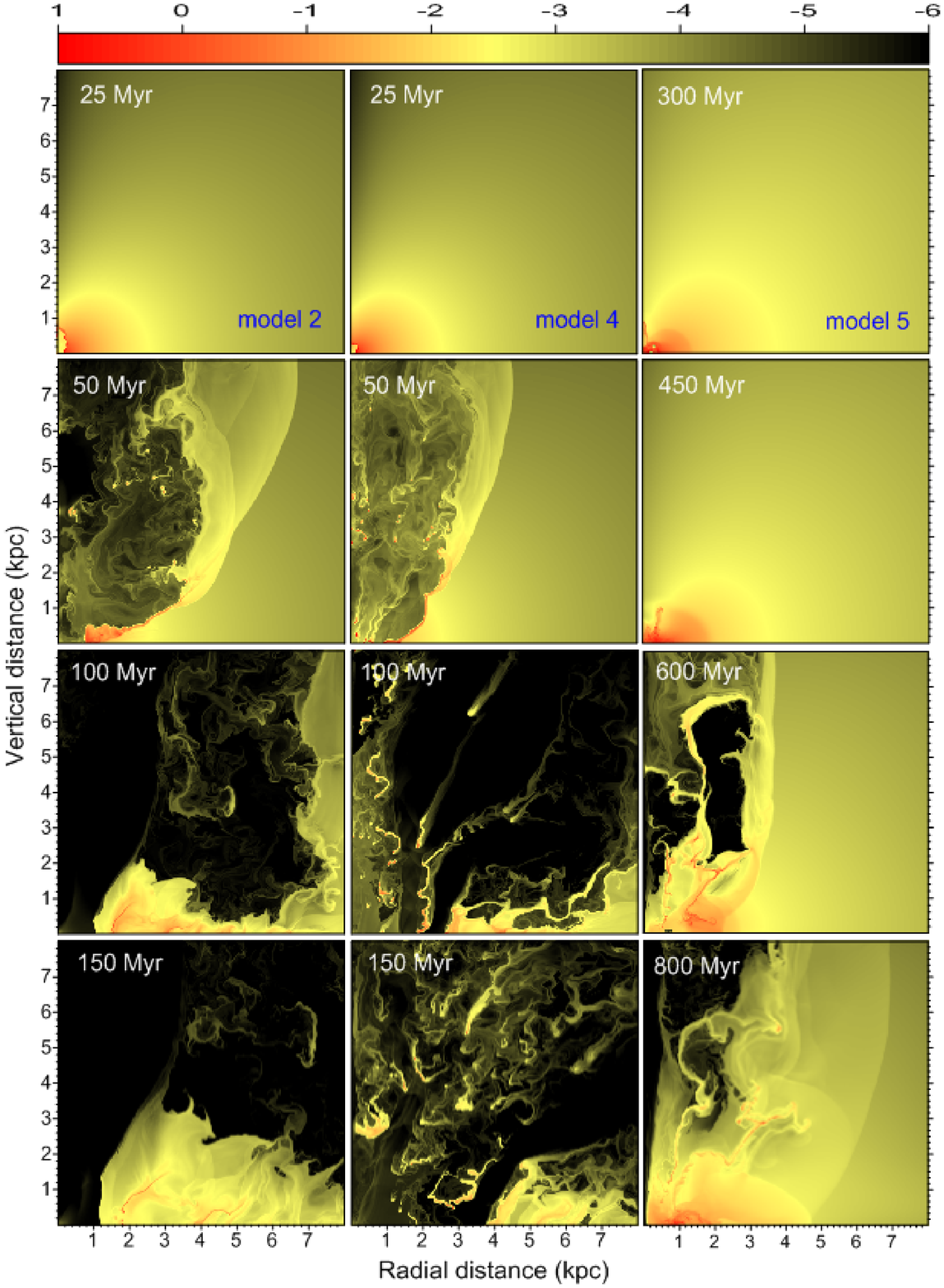}}
  \caption{Time evolution of the gas volume density in three models with $\alpha$=0.5.
    The l.h.s. column presents the time evolution in model~2, already shown in Fig.~\ref{fig3a}. 
    The middle
    column shows model~4, in which the motions of stars
    are artificially suppressed.  In the r.h.s. column we show model~5, 
    for which the initial equilibrium distribution was constructed without taking 
    gas self-gravity into account.
    Note the different evolutionary times in the latter model.}
         \label{fig8}
\end{figure*}

 In this work, we employ the moments of the collisional Boltzmann equation 
to describe the motion of stars. In addition, we start our simulations from 
self-consistent initial equilibrium conditions that were built taking gas 
self-gravity into account (in
some studies, the initial configuration is constructed
without taking gas self-gravity into account, although during the dynamical 
evolution the gas self-gravity may be included).
It is instructive to study how these two new recipes affect the
dynamics of our model galaxies.

In Figure~\ref{fig8} we show the gas volume density in three models with 
$\alpha$=0.5 at four representative evolution times.  In
particular, the left-hand-side column shows the evolution of our
reference model~2, already shown in Figure~\ref{fig3a}.  The middle column
shows model~4, in which the stellar dynamics has been artificially
suppressed.  Stars are allowed to form, but they are effectively
motionless and are fixed to their birthplaces until they die.
Evidently, the final gas distribution in model~4 is clumpy: the galactic volume is filled with moving
cometary-like clouds. The increase in the number of clumps is caused by a reduced feedback 
in the absence of stellar motion. The moving clumps leave behind the newly-born stars so that
part of the stellar energy is released  outside the clumps, which makes it easier for the clumps to
survive. In the presence of stellar motion, the newly-born 
stars follow the gas clumps (at least for some time) due to the fact that they inherit the gas velocity at the formation epoch, releasing the stellar energy inside the clumps and dispersing most of them.

The final gas distribution in model~2 is smooth and 
shows no cometary-like structures. At the same
time, the outflow in model~4 clears a larger
hole near the midplane than in the reference model. The combination of
these two effects results in a rather chaotic gas surface density
distribution shown in the middle panel of Figure~\ref{fig4} by the
dotted lines.  We also note that model~4 is
characterized by a somewhat higher star formation feedback, as is
indicated by the fraction of ejected gas mass shown by the
dot-dot-dashed line in Figure~\ref{fig5}.
 

The increased feedback in the model without stellar motions is likely
due to the fact that the energy of both supernova explosions and
stellar winds is released in regions characterized by a systematically
higher temperature and lower density than in the reference model. This
is because star formation creates regions of slightly lower density
compared to the surrounding gas, and stellar winds stir and heat up
the circumstellar gas.  If stars are motionless, then their feedback
is confined to their birthplaces.
On the other hand, if stellar motion is allowed, stars can move to
neighboring regions, where the density is likely to be higher and
where stellar winds have not yet heated the gas.  The feedback will be
thus less effective.  This conclusion is at variance with that of
\citet{Slyz}, who showed that stellar motion makes the feedback more
effective and can even solve the overcooling problem.  This different
outcome is probably due to the lack of feedback from stellar winds in
Slyz's models.

\begin{figure*}
  \hspace{2cm}\resizebox{13cm}{!}{\includegraphics{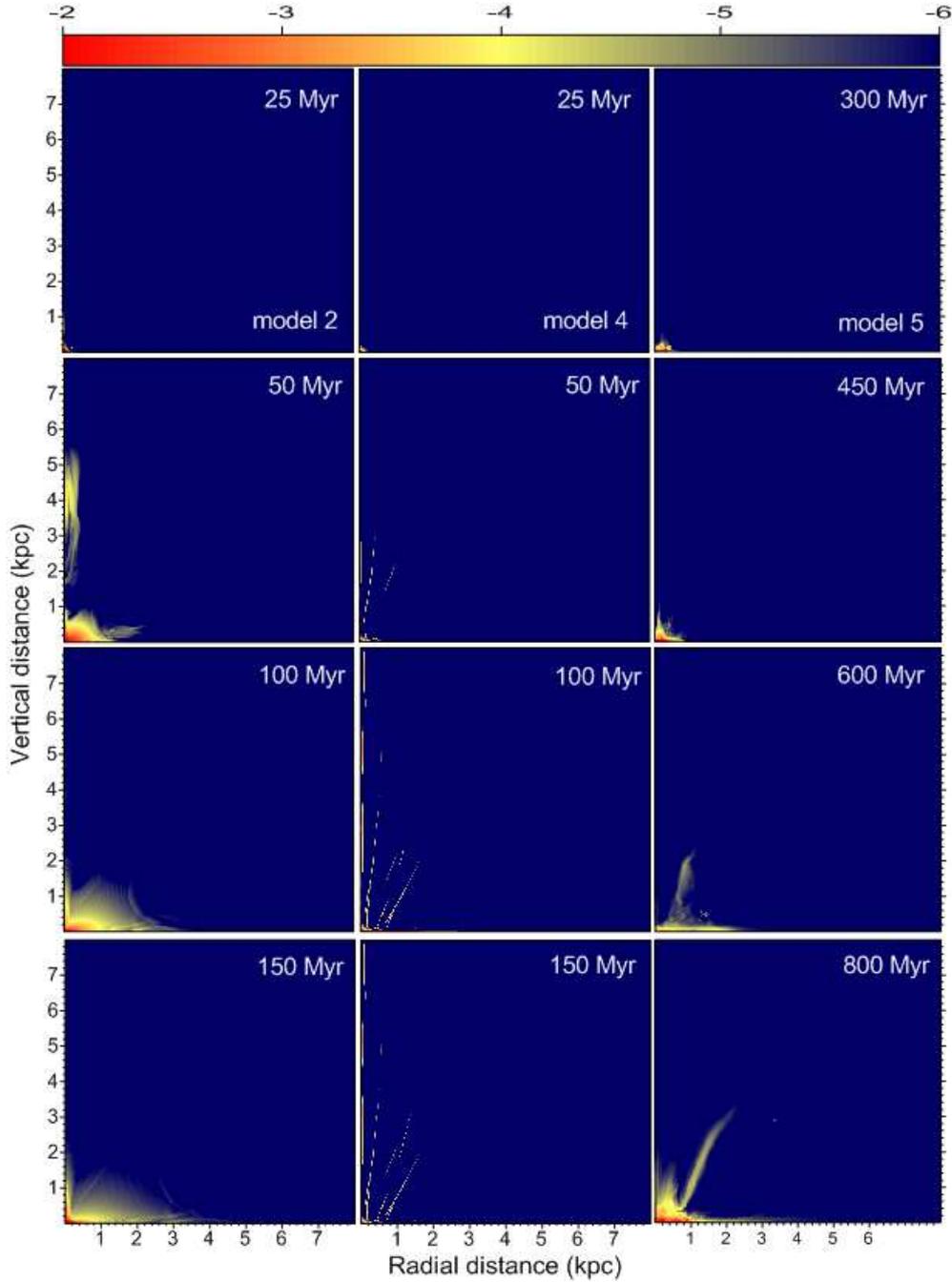}}
  \caption{Similar to Figure~\ref{fig8}, but now for the stellar
    volume density distribution.}
         \label{fig9}
\end{figure*}

Figure~\ref{fig9} presents the stellar volume density distributions 
for the same models as in Figure~\ref{fig8}. The left-hand column 
corresponds to the reference model~2, already shown in Figure~\ref{fig7}, 
while the middle column presents model~4, in which stellar motions are suppressed.
Evidently, models with and without stellar motion produce strikingly
different stellar distributions. The stellar distribution in the
reference model is rather smooth, whereas in the model without stellar
motions most of the stellar content is confined to the midplane and in
the inner region.  In fact, the fraction of the total stellar mass
confined within a vertical height of 0.3~kpc (from the midplane) is
82\% for the reference model, but this value increases to 98\% for the
model without stellar motions.  The curious stellar stripes visible in
the middle column are caused by the development of a gaseous
supershell: many stars tend to form on the walls of this supershell
and, as a consequence, their position reflects the shell's expansion.
The lack of stellar motion makes these structures permanent. The fact
that these structures are not observed in real galaxies outlines the
importance of a proper treatment of stellar dynamics in numerical
simulations of DGs.

Finally, we investigate the question of how the choice of the initial
gas density distribution can influence the subsequent evolution of
dwarf galaxies.  For this purpose, we choose $\alpha=0.5$ 
and construct two initial gas density distributions: one with the gas
self-gravity taken into account and the other without it.  
To make the comparison between
models with and without-self gravity easier, we require that the
gas mass within the 3-kpc radius be similar in both models.
The model with the initial equilibrium distribution taking gas self-gravity into
account is in fact our model~2 considered above, while its counterpart
is further referred as model~5 (see Table~\ref{table1} for model parameters).

The initial radial profiles of the gas surface density and the Toomre
$Q$-parameter in model~5 are shown in Figure~\ref{fig3} by the dash-dot-dotted lines.  
Not surprisingly, the initial gas distribution in model~2 is
more centrally condensed than in model~5: gas self-gravity creates an 
additional gravity pull towards the galactic center\footnote{The total gas 
and DM masses are comparable in the inner several kpc.}, resulting in a 
denser initial configuration.
For instance, the total gas mass in the inner sphere of 1~kpc in the
initial model without self-gravity is $\approx10^{7}~M_\odot$, while this
value rises by a factor of 2.8 in the model with self-gravity.  As a
consequence, the Toomre $Q$-parameter is initially greater than 2.0
everywhere and star formation is effectively suppressed.

The right-hand column in Figure~\ref{fig8} presents the gas
volume density distribution in model~5.   
We emphasize that the gas self-gravity in this model is neglected 
only when building the initial distribution; in the subsequent evolution 
the gas self-gravity is  however turned on again.
Evidently, the evolution of model~5 is notably slower owing to the fact 
that star formation is initially suppressed. The gas cools and contracts and only
after 120~Myr the conditions for star formation become favourable in
the galactic center, whereas a vigourous star formation feedback is
seen in model~2 already after 50~Myr of the evolution.  The 
initial slow start in model~5 is consistent with a long dynamical
timescale in this model. 
For the mean gas volume density in the inner 1~kpc 
of $4\times10^{-25}$~g~cm$^{-3}$, the corresponding free-fall time in model~5 is 
approximately 110~Myr, which is consistent with the time epoch of the first 
star formation episode in this model. Due to significant thermal support and radially increasing
dynamical timescale (density drops with radius), it takes several dynamical timescales
to adjust to a state similar to that of model~2. By the end of simulations in model~5 
(800~Myr), the outflow is still not as 
strong as in its counterpart model~2 and the central gap has not yet developed. 

The right-hand column in Figure~\ref{fig9} presents the stellar volume density 
distribution in model~5. A visual inspection of the l.h.s. and r.h.s. models reveals that 
the stellar distributions in models~2 and 5 are in general similar
(but note the difference in ages 150 vs. 800 Myr), 
though the stars in the latter model are distributed somewhat more irregularly. 
By the end of numerical simulations, the total mass of
formed stars in the 800-Myr-old model~5  is still a factor of 4 lower than in the 150~Myr-old model~2,
highlighting a much slower evolution in the model that starts from a 
non-self-consistent initial configuration, i.e. without initial gas 
self-gravity.


\section{Discussion and model caveats}
\label{caveats} 

In this work, we did not aim at a detailed
  comparison of our numerical simulations with observed DGs, which is
  our task for subsequent studies.  Instead, we have chosen three
  typical models for DGs with different rotational support against
  gravity to test the code performance on global simulations, and not
  just on a predefined set of classic test problems. The current
  version of the code assumes axial symmetry and neglected chemical reactions,
   which leaves us a wide door for future improvements. At
  the same time, we have managed to develop an efficient mathematical
  recipe for describing the phase transition from stars into gas using
  the continuity equation for the mean stellar age (see Section~2.6).

As already mentioned in the text, some simulation parameters
have been chosen without a detailed parameter study or without an
in-depth analysis.  We avoid such a detailed study
for the sake of clarity and in order to make the paper more concise
and readable.  A deeper parameter study will be performed in
follow-up papers.  It is nevertheless important to point out the
 potential caveats in the choice of some parameters.

\begin{itemize}

\item {
{\it $\epsilon_{\rm sw}$ (see Sect. 2.7.3.):}
Massive and intermediate-mass stars expel a significant
fraction of their mass by stellar winds. Although their 
lifetimes shorten with mass, the mass-loss rates increase
disproportionately, so that for solar abundances the total wind
energy of most massive stars exceeds that of SNeII over the 
stellar lifetime.  
For radiation-driven HII regions, \citet{Lasker67}
stated that the ISM is heated only by a very low amount and
derived analytically an efficiency of 1\%, because the ionizing 
radiation energy is almost instantaneously lost by line emission
and only the expansion of HII regions steers up the surrounding ISM.
Although the additional wind power amounts to almost 10\% of the 
radiation energy, it is expected to deposit a larger fraction as
the thermal and kinetic ISM energies. 
Since in most chemo-dynamical simulations \citep{Theis92,Samland}
the Ly continuum radiation is applied with an efficiency of 10$^{-3}$,
$\epsilon_{\rm sw}$ is assumed to 100\%. This value is also used here.

\item
{\it $\epsilon_{\rm SN}$ (see Sect. \ref{SNtwo}):} 
For the thermalization efficiency of the SNII explosion, i.e. the 
fraction of the explosion energy transferred to gas thermal energy we 
have adopted the value $\epsilon_{\rm SN}=1.0$ for simplicity
and in order to avoid further parameter studies. The thermalization 
efficiency has been subject to considerable studies in the past. 
\citet{Thorn98} performed single 1D supernova explosion models 
and derived $\epsilon_{\rm SN}=0.1$. Estimates from cumulative
SNII explosions, forming superbubbles, suggest a significant
dependence of $\epsilon_{\rm SN}$ on the temperature and density of
the external medium \citep{Recchi01, recchi14}.  }
A different approach, where the contribution of a whole stellar
population is considered \citep{Melio04} clearly shows that
$\epsilon_{\rm SN}$ is a function of time.
 We have performed a test simulation adopting $\epsilon_{\rm SN}=0.1$. 
 The extent of the galactic winds was reduced in this
  case, but not by a large factor.  It is in fact important to remark
  that with our formulation for the SFR (see Eq. \ref{sfr}), a great
  deal of self-regulation is achieved.  In fact, the larger
  $\epsilon_{\rm SN}$, the larger the heating of the surrounding
  medium, the higher the probability that the star formation is
  quenched for a period of time \citep[see][for more details on
  self-regulated star formation]{KTH95}.  Thus, although we plan to
  make more careful choices of $\epsilon_{\rm SN}$ in follow-up
  papers, this parameter appears not to change drastically the results
  of the simulations.

\item 
$n_{\rm cr}$ (see Sect. \ref{subs:starformationrate}). The 
threshold density above which star formation is allowed, $n_{\rm cr}=1$~cm$^{-3}$ has been
  chosen in compliance with other numerical studies
  \citep{Springel2003, Recchi07}.  Also in this case, many different
  values for $n_{\rm cr}$ have been chosen in the literature and
  detailed studies on the effect of this parameters have been
  performed.  In the framework of $\Lambda$CDM cosmological
  simulations,  $n_{\rm cr}$ of 0.1 cm$^{-3}$ is generally
used \citep{Katz1992} and its effect on the star formation was studied, e.g.,
by \citet{Kay2002}, but also applied to much larger resolutions of
smooth-particle hydrodynamics applications \citep{Stin06}. 
  It appears that an extremely large value of 
\citep[up to 100 cm$^{-3}$][]{Gove10} is required to obtain DGs with
  structural properties similar to the observed ones \citep[but see
  also][]{Pilk11}.  Quite generally, the star formation density
  threshold should be set based on the density at which gravitational
  instabilities can be resolved \citep[see e.g.][]{True97}.  This leads
  for instance \citet{Stin13} to assume a threshold of 9.3 cm$^{-3}$
  based on their adopted resolution.  Clearly, our choice of $n_{\rm
    cr}$ is simplistic and deeper studies are required in follow-up
  publications.  We notice only that in our simulation only a tiny
  fraction of gas reaches densities larger than a few tens of
  cm$^{-3}$, therefore the adoption of very large values of $n_{\rm
    cr}$ would lead to extremely localized and sporadic star
  formation.  We notice also that the very large feedback achieved in
  the simulations of \citet{Gove10} is certainly able to produce
  modelled DGs with structural properties similar to the observed
  ones, but it is not clear whether the chemical properties can be
  equally well reproduced.  A very large feedback can create galactic
  winds, able to expel the large majority of metals, freshly produced
  in the galaxy, thus at the end of the evolution the model galaxy
  should be almost deprived of heavy elements.

\item $\Lambda_{\rm h}$ (see Sect. \ref{subs:overcooling}).  The cooling 
function typical of hot gas $\Lambda_{\rm h}$, which has been used to
  calculate the cooling timescale of the hot SN ejecta, has been
  estimated based on a temperature of the hot ejecta of 10$^8$ K.  It
   must be admitted that this temperature only holds just after the 
SN explosion, but the ejecta cools significantly afterwards. 
 Nevertheless, the thermal SN energy has been varied artificially 
to probe its influence on galactic winds \citep{DVS2012}. 
We note, however, that we are in the cooling regime dominated by
  bremsstrahlung, and the temperature dependence of the cooling
  function is not so severe (the cooling is approximately proportional
  to $\sqrt{T}$ in this temperature range).
\end{itemize}

\section{Conclusions}
\label{conclusions}
In this paper, we presented the stellar hydrodynamics approach for
modeling the evolution of DGs, paying special
attention to the co-evolution of gas and stars. The distinctive
feature of our models is how we treat the stellar component. Unlike
many previous studies that use various versions of the N-body method
to compute the dynamics of stars, we employ the stellar hydrodynamics
approach based on the moments of the collisionless Boltzmann equation,
originally introduced by \citet{Burkert88}, and applied and further developed by 
\citet{Theis92,Samland,VT06,VT08,Mitchell,Kulikov2014}. 
For the first time, 
we provided an effective mathematical recipe for describing the death of stars 
in the framework of stellar hydrodynamics using the concept of a mean stellar age. We performed an
extensive testing of our numerical scheme using classical and newly
developed test problems, paying particular attention to the proper
recipe for artificial viscosity in order to achieve the best agreement
with the analytical solution for the Sedov test problem.

We demonstrate the success of our numerical approach by simulating the
evolution of three DGs differing in the amount of rotation described
by the spin parameter $\alpha$=0.25, $\alpha$=0.5, and $\alpha$=0.9.
In agreement with previous studies, models with low values of $\alpha$
are characterized by stronger outflows, ejecting a larger amount of
gas from the galaxy.  This is due to the fact that the initial gas
distribution in the low-$\alpha$ models is almost spherical, so that
the gas pushed by the stellar feedback expands almost isotropically.
On the contrary, models with high values of $\alpha$ exhibit a
disk-like initial gas distribution with a pronounced radial flaring.
The pressure gradient in the vertical direction in these models is
very steep, which favours the development of bipolar galactic
outflows.  The transport of gas in the horizontal direction (i.e.
along the galactic midplane) is very limited, and therefore this model
is able to keep most of the initial gas bound to the galaxy.  The
fraction of ejected gas (i.e., the fraction of gas that leaves the
computational domain with a speed higher than the escape speed) is
only a few percent for the $\alpha=0.9$ model, but it raises to 80\%
for the model with $\alpha=0.25$.

At the same time, models with a low value of $\alpha$ produce a
more diffuse stellar population than models with higher $\alpha$. Only
models with $\alpha\ge 0.5$ reveal the formation of a definite stellar
disk component.  We demonstrate the importance of the stellar dynamics
by artificially turning off stellar motions. In this case, the
resulting stellar population is mostly concentrated to the midplane
and shows stripe-like structures emanating from the galactic center,
which are not observed in real galaxies.  

Stars start heating the natal gas almost instantaneously
owing to the effect of stellar winds neglected in many recent hydrodynamical
simulations.  We found that if
stellar dynamics is suppressed, Type II supernovae (the main source
of mechanical feedback) explode in a medium heated and diluted by the stellar winds.  
In such an environment, radiative losses are small and
the supernova feedback is extremely effective.  If, on the other hand, 
stellar motions are allowed,
they can move from the natal site (where the gas has been heated and
stirred by the effect of stellar winds) to neighboring regions
before exploding as supernovae.  These regions are characterized by
lower temperatures and higher densities, so that the supernova feedback 
becomes less effective.  It is thus crucial to correctly simulate stellar
dynamics (together with all relevant sources of feedback) in order
to study the interplay between stars and the interstellar medium.

Unlike many other studies of DGs, we used the self-consistent initial gas
configuration that was constructed taking the gas self-gravity into
account \citep{VRH1}. This gas configuration is usually more centrally condensed
and prone to star formation than that constructed neglecting gas self-gravity.  
We compared the evolution of our model galaxy starting from these different initial
configurations  and found that models that start from non-self-gravitating
initial configurations evolve much slower than those that start from 
self-gravitating ones. In the former, a major episode  of star formation and the development
of prominent outflows may by delayed by 0.5-0.8~Gyr as compared to the latter.
We note that this is exclusively the effect of initial conditions (or, better to say, the effect of
using non-self-consistent initial gas configurations), because 
the gas self-gravity is turned on in both models once the evolution has started.

In the future, a detailed comparison of our stellar hydrodynamics
simulations with a more conventional N-body treatment of stellar
dynamics is desirable to assess the weak and strong sides of the
Boltzmann moment equation approach.  This project was partly supported
by the Russian Ministry of Education and Science Grant  3.961.2014/K 
and a Lise Meitner Fellowship, project
number M 1255-N16. The simulations were performed on the Vienna
Scientific Cluster (VSC-2). This publication is supported by the
Austrian Science Fund (FWF). We are thankful to the anonymous 
referee for a very insightful review which helped to  
improve the manuscript.



\appendix
\section{Testing gas hydrodynamics equations} 
\label{gastest}
Our numerical scheme for solving the equations of gas
hydrodynamics~(\ref{contingas})--(\ref{energy}) can be tested using a
conventional set of test problems suitable for cylindrical
coordinates.  Below, we provide the results of six 
 essential test problems, each designed to test the code
performance at various physical circumstances.  The code has also
performed well on two additional tests: the two interacting blast
waves and the Noh (or implosion) problems, but we do not provide the
results for the sake of conciseness.

\subsection{Sod shock-tube test}
\label{SodGas}
This test is often used to assess the ability of a numerical algorithm
to accurately track the position of relatively weak shock waves and
contact discontinuities.  Initial conditions involve two discontinuous
states along the z-axis, with a hot dense gas on the left and cold
rarified gas on the right. More specifically, we set the pressure and
density at $z\in [0-0.5]$ to 1.0, while at $z\in [0.5-1.0]$ the
pressure is 0.1 and density is 0.125. The velocity of a $\gamma=1.4$
gas is initially zero everywhere.

\begin{figure}
  \resizebox{\hsize}{!}{\includegraphics{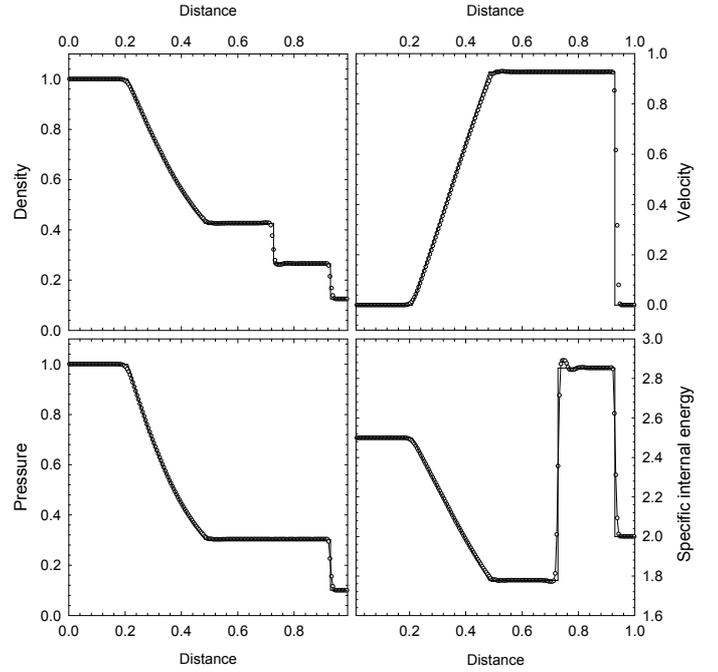}}
  \caption{Sod shock tube problem for the gas density (upper-left),
    velocity (upper-right), pressure (lower-left), and specific
    internal energy (lower-right).  The numerical solution is shown by
    open circles, while the analytical one is plotted by solid lines.}
         \label{fig1ap}
\end{figure}

Filled circles in Figure~\ref{fig1ap} show the results of the Sod
shock-tube test computed with a resolution of 200 grid zones and
$C_2=4$ (the number of zones over which the shock is spread), while
the solid lines plot the analytic solution at $t=0.235$. It is evident
that the code tracks well the positions of shock waves and contact
discontinuities, which are resolved by 3--4 zones, in accordance with
the chosen value of $C_2$.  Decreasing $C_2$ results in sharper shocks
and contact discontinuities but may increase the magnitude of
anomalous spikes in the specific energy (bottom-right panel), and
hence not recommended.  Increasing $C_2$ to 6 produces results
similar to those for $C_2=4$, except that the shocks and contact discontinuities
are now resolved by 5-6 zones. 

\subsection{Sedov point explosion}
\label{sedov}
The Sedov explosion problem tests the code's ability to deal with
strong shocks. In contrast to the Sod shock-tube test, in which shock
waves were of a pure planar geometry, this test involves a spherically
symmetric strong shock wave.  Hence, the point explosion is a powerful
test on the code's ability to render spherically symmetric structures
on an essentially non-spherical grid stencil.

\begin{figure}
  \resizebox{\hsize}{!}{\includegraphics{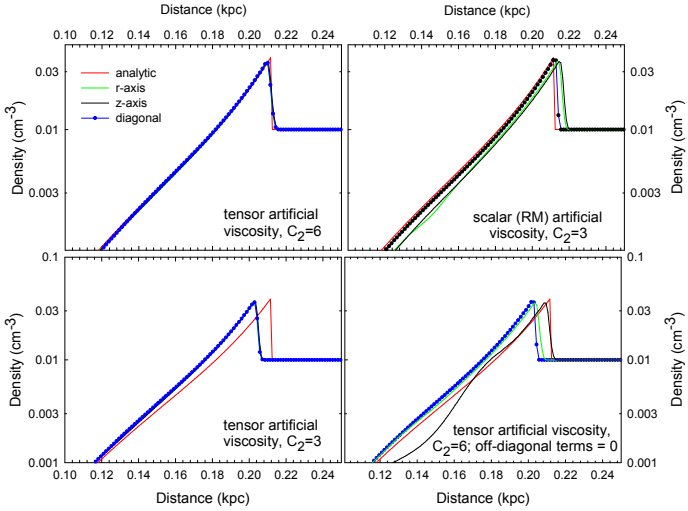}}
  \caption{Sedov point explosion test for the explosion energy
    $10^{51}$ erg and background density $n=0.01$~cm$^{-3}$. The
    analytic solution is shown by the red lines, while the numerical
    solutions along the $z$-axis, $r$-axis, and diagonal $z=r$ are
    shown by the black, pink, and blue lines, respectively. The
    top-left panel presents the solution for the tensor artificial
    viscosity with $C_2=6$, the top-right panel---for the scalar
    artificial viscosity and $C_2$=3, the bottom-left panel---for the
    tensor artificial viscosity and $C_2=3$, and the bottom-right
    panel---for the tensor artificial viscosity with zero off-diagonal
    terms and $C_2$=6.}
         \label{fig2ap}
\end{figure}

To initialize the explosion, we consider a cold ($T=10$~K) homogeneous
medium with number density $n=10^{-2}$~cm$^{-3}$ and inject $10^{51}$
ergs of thermal energy (equivalent to one supernova) into the
innermost grid cell.  The adopted resolution is $300 \times 300$ grid
cells and the size of each cell is 1.0~pc in each coordinate direction
($z$ and $r$).  We emphasize that we do not inject energy into a
central sphere comprising a few cells near the coordinate origin---a
usual practice to alleviate the problem of a non-spherical geometry.
Instead, we consider the most difficult situation when energy is
injected in just one, essentially cylindrical innermost cell.

Figure~\ref{fig2ap} compares our derived density distributions along
the z-axis (black line), the r-axis (pink line), and the diagonal
$z=r$ (blue line with circles) with the analytic solution given by the
red line \citep{Sedov} at $t$=1~Myr after the explosion. In
particular, numerical simulations in the top-left panel are obtained
using the tensor artificial viscosity (AV) as defined in
Section~\ref{solution}, while the density distribution in the
top-right panel is obtained using the ''classic'' scalar AV as
described by \citet{RM}.  It is evident that the tensor AV yields a
much better agreement with the analytic solution. The expanding shell
has almost a perfect spherical shape (all three lines showing the
numerical solution virtually coincide), notwithstanding the fact that
the energy was injected into a cylindrically shaped innermost cell. On
the other hand, the Richtmyer \& Morton scalar AV yields a notable
deviation from sphericity along the $r$ and $z$ axes, as seen in the
top-right panel of Figure~\ref{fig2ap}. The Richtmyer \& Morton solution
also slightly overshoots the analytical one.  Regardless of the AV
prescription used, the shock front is resolved by 2--3 grid zones.

We want to emphasize that all components of the artificial viscosity
tensor $\bf Q$ should be used, including the off-diagonal terms.
Neglecting the latter, as is done in, e.g., \citet{SN92}, leads to a
significant deterioration of the numerical solution.  The bottom-right
panel in Figure~\ref{fig2ap} demonstrates the effect of zeroing the
off-diagonal terms in $\bf Q$. As one can see, the numerical solution
is skewed, notably undershooting the position of the shock front in the
$r$-direction and along the diagonal direction but reproducing the shock front rather well
in the $z$-direction.

\begin{figure}
  \resizebox{\hsize}{!}{\includegraphics{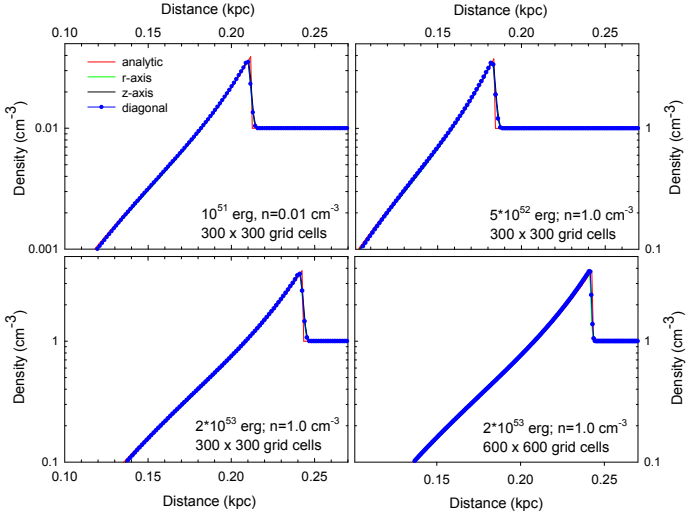}}
  \caption{Sedov point explosion test for various explosion energies,
    background densities and numerical resolutions as indicated in
    each panel. The tensor artificial viscosity with $C_2=6$ is used
    for every plot. The line meaning is the same as in
    Figure~\ref{fig2ap}.}
         \label{fig3ap}
\end{figure}

A caution should be paid when choosing the AV softener $C_2$.  As one
can notice in Figure~\ref{fig2ap}, the test was done with different
values of $C_2$ for the tensor and scalar AV, 6 and 3, respectively.
If we choose $C_2=3$ for the tensor AV, the position of the shock
front notably undershoots the analytic solution, as is illustrated in
the bottom-left panel of Fig.~\ref{fig2ap}. Fortunately, once the
proper value of $C_2$ is found, the tensor AV algorithm performs well
regardless of the energy input, background density, or numerical
resolution. We demonstrate this in Figure~\ref{fig3ap} showing the
tensor AV performance for various choices of the input parameters as
indicated in each panel and $C_2=6$.  It is evident that the test
results are virtually insensitive to a particular choice of initial
conditions.

The sensitivity of the Sedov test to a particular choice of the AV
softener $C_2$ is a well known problem of numerical codes that evolve
the internal energy \citep[e.g.][]{Tasker}. Switching to the total
energy helps to eliminate this problem \citep[e.g.][]{Clarke10}.
However, in numerical codes that make use of a staggered grid as our
own, with vector and scalar variables defined at different positions
on the grid, solving for the total energy equation is not a good
option as it inevitably requires interpolating between the internal
and kinetic energies to obtain the total one \citep{Clarke10} or
applying corrections after each time step to conserve the total energy
\citep{Vshivkov11}. In either case, this practice, according to our
experience, often comes at a cost of degraded performance on other
test problems and is not recommended. Instead, we show that a proper
choice of the artificial viscosity softener can help to resolve this
problem without resorting to the total energy equation.



\subsection{Collapse of a pressure-free sphere}
\label{collapse}
The gravitational collapse of a pressure-free sphere is used to assess
the code's ability to accurately treat converging spherical flows in
cylindrical coordinates.  This test is also useful for estimating the
performance of the gravitational potential solver on dynamical
problems (our solver does well on the static configurations such as
spheres and ellipsoids).  To run this test, we set a cold homogeneous
sphere of unit radius and density (for convenience, the gravitational
constant is also set to unity) and let it collapse under its own
gravity. The analytic solution to this problem exists only in the
limit of an infinite cloud radius, describing the collapse of every
mass shell \citep{Hunter62}.  In the cylindrical geometry, however, we
consider a cloud of finite radius with a sharp outer boundary to
preserve the cloud sphericity.  As a result, a rarefaction wave
develops after the onset of the collapse, propagating towards the
coordinate origin and necessitating complicated corrections to the
analytic solution \citep{Truelove98}.

\begin{figure}
  \resizebox{\hsize}{!}{\includegraphics{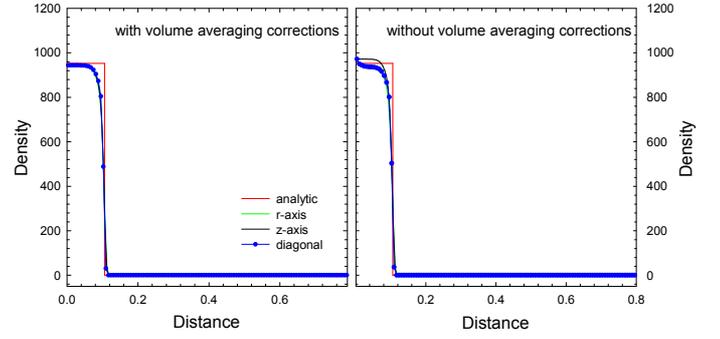}}
  \caption{Collapse of a pressure-free sphere showing the gas density
    at 0.985 that of the free-fall time. The left/right panels show
    the test results with/without the volume averaging corrections as
    described in \citet{Blondin93}. The line meaning is the same as
    in Figure~\ref{fig2ap}.}
         \label{fig4ap}
\end{figure}

The left panel in Figure~\ref{fig4ap} compares the results of our
numerical simulations with the ''uncorrected'' analytic solution of
\citet{Hunter62} (red line) at $t$=0.535 (or 0.985 that of the
free-fall time) when the initial density has increased by nearly three
orders of magnitude. As in the previous test, we plot the numerically
derived density along the $z$-axis (black line), $r$-axis (green
line), and the $z=r$ diagonal (blue line with circles). The numerical
resolution is $200\times200$ grid zones. It is seen that the cloud's
outer boundary is somewhat smeared out due to the action of the
rarefaction wave. The peak density is also about $1\%$ smaller than
that predicted from the analytic solution. On the other hand, the code
performs well on preserving sphericity of the collapsing cloud. In the
right panel, we turn off the volume averaging corrections applied to
the advection of the $r$-momentum in the $r$-direction (as described
in \citet{Blondin93}), which results in the development of an
artificial density spike near the $z$-axis and also in the distortion
of a spherically symmetric collapse. This test demonstrates the
utility of the volume averaging corrections introduced originally by
\citet{Blondin93} in the PPA advection scheme in non-Cartesian
coordinates.

\subsection{Collapse of a rotating sphere}
\label{rotation}
This test is invoked to assess the code's ability to conserve angular
momentum.  The setup consists of a isothermal (T=10~K) homogeneous
sphere of unit radius and density, which rotates at a constant angular
velocity $\Omega_0$. The latter is found from the requirement that the
ratio $\beta$ of the rotational energy $E_{\rm rot}=I \Omega^2_0/2$ to
gravitational energy $E_{\rm grav}=3 G M_0^2/(5 R)$ is equal to $5\%$.
Here, $I = 2 M_0 R^2/5$ is the moment of inertia of a homogeneous
sphere of radius $R$ and mass $M_0$. The adopted value of $\beta=0.05$
is close to an upper limit found in rotating pre-stellar molecular
cloud cores and dark matter halos.

In the absence of any mechanisms for angular momentum redistribution,
the mass with specific angular momentum less than or equal to
$K=\omega r^2$, where $\omega$ is the angular velocity at a distance
$r$ from the $z$-axis, should be conserved and equal to
\citep{Norman80}
\begin{equation}
\label{angmom}
M(K)=M_0 \left[ 1 - \left(1- {K \over \Omega_0 R^2 } \right)^{3/2} \right].
\end{equation}
A deviation from equation~(\ref{angmom}) would manifest the
non-conservation of angular momentum in the numerical algorithm.

\begin{figure}
  \resizebox{\hsize}{!}{\includegraphics{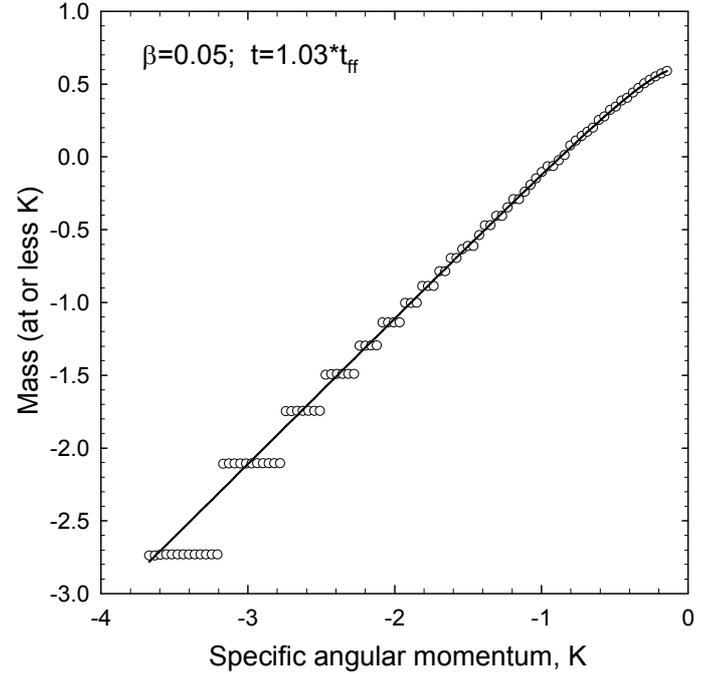}}
  \caption{Specific angular momentum spectrum M(K), calculated using
    equation~(\ref{angmom}), for a collapsing sphere at $1.03 t_{\rm
      ff}$. The analytic spectrum is shown by the solid line, while
    the test results are plotted by open circles.}
         \label{fig5ap}
\end{figure}

The solid line in Figure \ref{fig5ap} shows the initial mass spectrum
$M(K)$, whereas the open circles present the mass spectrum obtained at
time $1.03\times t_{\rm ff}=0.551$, where $t_{\rm ff}$ is the
free-fall time for a non-rotating sphere of the same density and
radius.  The numerical resolution is $200\times200$ grid zones.  By
$t=0.551$, the density near the center of the sphere has increased by
more than three orders of magnitude, but the deviation of the mass
spectrum from the initial configuration is negligible everywhere
except at lowest values of $K$. This deviation is a manifestation of
angular momentum non-conservation near the rotational axis caused by
imperfections in the interpolation procedure across the inner
$r-$boundary.  Nevertheless, the total mass of gas that suffers form
the angular momentum non-conservation is just below $1\%$ and
therefore should not affect notably the global evolution.


\subsection{Kelvin-Helmholtz Instability}
The Kelvin-Helmholtz instability (KHI) occurs at the interface between
two fluid moving with different velocities. In the context of this
paper, this instability operates at the interface between the hot
supernova ejecta and the cold swept-up shell and may cause an
accelerated disintegration of the latter \citep[e.g.][]{VB05}.
Therefore, it is essential that the code be able to demonstrate the
development of the KHI as expected from the linear perturbation
theory.

The initial setup involves two fluids moving in opposite directions
along the $z$-axis with the relative velocity $v_{\rm z,rel}=0.2
c_{\rm s}$, where $c_{\rm s}=0.06$ is the sound speed.  The interface
between the fluids is located at $r=0.5$ and the densities of the
inner and outer fluids are $\rho_{\rm in}=1.0$ and $\rho_{\rm
  out}$=0.1, respectively.
We use periodic boundary conditions and the resolution is $200\times
200$ grid zones on the $1.0\times1.0$ computational domain.

To trigger the instability, we impose a sinusoidal perturbation on the
radial velocity $v_{\rm r}$ of the form
\begin{equation}
v_{\rm r}=\delta v_{\rm r} \sin(2\pi z/\lambda),
\end{equation}
where the amplitude and wavelength of the perturbation are $\delta
v_{\rm r}=v_{\rm z,rel}/100$ and $\lambda=1/6$, respectively. The
perturbation is applied to five neighboring grid zones on both sides
of the fluid interface.

\begin{figure}
  \resizebox{\hsize}{!}{\includegraphics{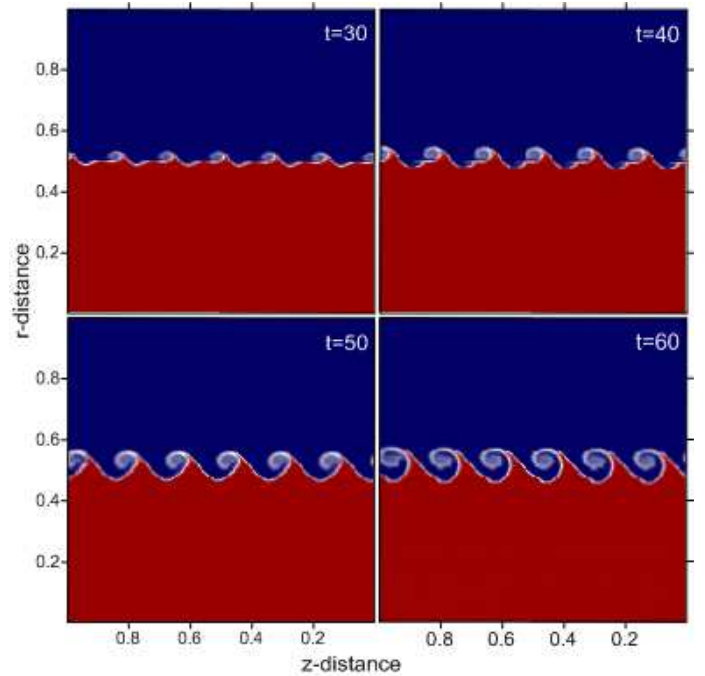}}
  \caption{Development of the KHI at different moments in
      time (see text for more detail).}
         \label{fig6ap}
\end{figure}

Figure~\ref{fig6ap} demonstrates the growth of the KHI with time.
There are six growing vortices, in agreement with the wavelength of
the perturbation $\lambda=1/6$, and the growth time is approximately
50--60, again in agreement with the characteristic growth time of the
KHI for our choice of the fluid density and velocity
\begin{equation}
\tau_{\rm KH}={ \lambda (\rho_{\rm in}+\rho_{\rm out}) \over  v_{\rm
z,rel} \sqrt{\rho_{\rm in} \rho_{\rm out}} }=48.3.
\end{equation}
\noindent 
We should note here that, for this particular test, gravity can be
neglected.  Therefore, all modes of perturbation, irrespective of
their wavelenghts (or wavenumbers) are unstable \citep[see][for
stability criteria when gravity is taken into account]{VH07}.

\subsection{Overstability of a radiative shock}
A useful test problem for hydrodynamic codes with optically thin
radiative cooling is described in \citet{SN93}. The test involves
setting a unform flow of gas with velocity $v_0$ against a reflecting
wall. An adiabatic reflection shocks forms immediately at the wall and
propagates upstream with velocity $v_{\rm sh}=v_0/3$. After
approximately one cooling time, the shock looses its pressure support
and is then advected back to the wall at $v_{\rm sh}\la-v_0$.  After
reflecting off the wall, the shock again becomes adiabatic and
propagates outward to repeat the cycle.

\begin{figure}
  \resizebox{\hsize}{!}{\includegraphics{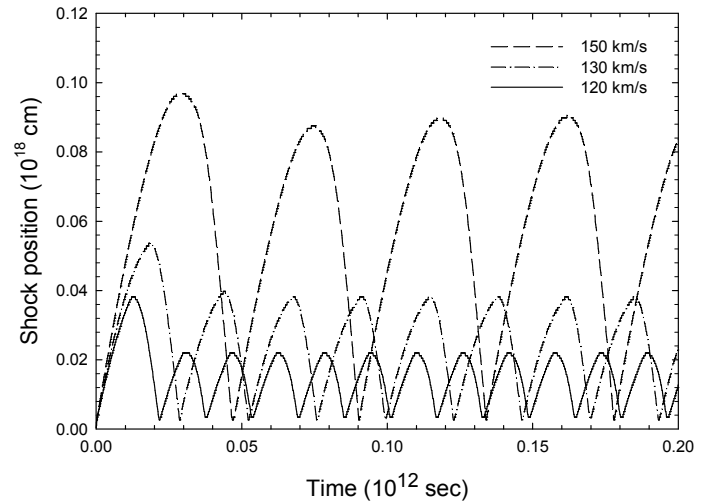}}
  \caption{Shock position versus time of a one-dimensional radiative
    shock during overstable oscillations. The shock positions for the
    flow velocities of 120~km~s$^{-1}$, 130~km~s$^{-1}$ and
    150~km~s$^{-1}$ are plotted with solid, dash-dotted, and dotted
    lines, respectively.}
         \label{fig7ap}
\end{figure}

Figure~\ref{fig7ap} presents the test results for the cooling function
with solar metallicity and ionization fraction $f_{\rm ion}=10^{-3}$.
Gas heating is turned off for this test.  The initial setup is
identical to that described in \citet{SN93}, i.e., the initial gas
number density and temperature are set to 10~cm$^{-3}$ and $10^4$~K,
respectively.  The position of the shock front for three flow
velocities of 120~km~s$^{-1}$, 130~km~s$^{-1}$ and 150~km~s$^{-1}$ are
plotted with the solid, dash-dotted, and dashed lines, respectively.
The shock position demonstrates regular pulses, as expected. The
amplitude and period of the observed oscillations are somewhat
different from those presented in \citet{SN93}, which simply reflect
the difference in the adopted cooling functions. The upstream velocity
of the adiabatic shock is close to the expected value, $v_0/3$. For
instance, in the case of $v_0=150$~km~s$^{-1}$, the maximum upstream
velocity of the shock is 47~km~s$^{-1}$. This test demonstrates the
reliability of our solution scheme for the gas cooling and heating
described in Section~\ref{coolheat}.

\section{Testing stellar hydrodynamics equations}

Testing the stellar hydrodynamics
equations~(\ref{continstar})--(\ref{dispstar}) presents a certain
challenge since stellar systems are described by a stellar dispersion
tensor, which may be anisotropic, rather than by an isotropic
pressure.  Fortunately, some of the test problems considered above can
be adapted to stellar hydrodynamics as well. The matter is that the
equations of stellar hydrodynamics can be formally made identical to
those of gas hydrodynamics for a specific set of initial conditions
(note that the continuity equations are always identical).  Indeed,
for a one dimensional flow along the $z$-direction (with zero gravity
and no star formation), equations~(\ref{momstar}) and (\ref{dispstar})
become
\begin{equation}
{\partial \over \partial t} \left( \rho_{\rm s} v_{{\rm s},z} \right) + 
{\partial \over \partial z} \left( \rho_{\rm s} v_{{\rm s},z} \cdot v_{{\rm s},z} \right) + 
{\partial \over \partial z} \left( \rho_{\rm s} \sigma_{zz}^2 \right) =0,
\end{equation}

\begin{equation}
{\partial \over \partial t} \left( \rho_{\rm s} \sigma_{zz}^2 \right) +
{\partial \over \partial z} \left( \rho_{\rm s} \sigma_{zz}^2 \cdot v_{{\rm s},z} \right) + 
2\rho_{\rm s} \sigma_{zz}^2 {\partial v_{{\rm s},z} \over \partial z}=0.
\end{equation}
These equations become identical to the corresponding gas dynamics
equations for $\rho_{\rm g} v_{z}$ and $\epsilon$, if we set
$\rho_{\rm s} \sigma_{zz}^2=P$, $\epsilon=(\gamma -1) \rho_{\rm s}
\sigma_{zz}^2$, and $\gamma=3$ \citep[see][for details]{Mitchell}.
This fact allows us to directly compare the performance of both
stellar and gas hydrodynamics code on some test problems considered in
Section~\ref{gastest} or use analytic solutions available from the gas
dynamics.


\subsection{Sod shock-tube test}
The initial setup for this test is identical to that considered in
Secion~\ref{SodGas}.  We run the test along the $z$-axis and set
$\rho_{\rm s} \sigma_{zz}^2$ and $\rho_{\rm s}$ at $z\in [0-0.5]$ to
1.0, while at $z\in [0.5-1.0]$ the z-component of the stress tensor is
0.1 and stellar density is 0.125.

\begin{figure}
  \resizebox{\hsize}{!}{\includegraphics{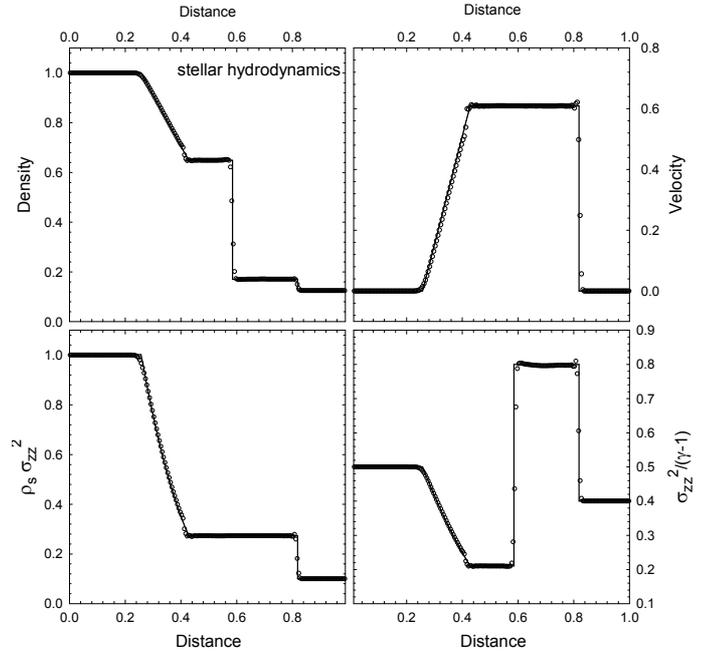}}
  \caption{Sod shock tube problem for the stellar density
    (upper-left), velocity (upper-right), $z$-component of the stress
    tensor (lower-left), and square of the $z$-velocity dispersion
    (lower-right).  The numerical solution is shown by open circles,
    while the analytical one is plotted by solid lines.}
         \label{fig8ap}
\end{figure}

Our numerical results for $C_{2s}=4$ and $t=0.14$ (open circles) are
compared against the analytical solution (solid lines) in
Figure~\ref{fig8ap}. It is seen that our numerical scheme correctly
reproduces the position of shock waves and contact discontinuities in
the stellar fluid.  On the other hand, small-scale oscillations are
evident at shock positions, perhaps reflecting the lack of energy
dissipation due to the collisionless nature of stellar hydrodynamics.
An increase in the coefficient of artificial viscosity helps to reduce
the amplitude of spurious oscillations but simultaneously increases
the shock smearing and hence is recommended only if these oscillations
lead to numerical instabilities.

\subsection{Collapse of a cold sphere and angular momentum conservation}
The collapse of a cold stellar sphere, both with and without rotation,
have produces very similar results to those of presented in
Sections~\ref{collapse} and \ref{rotation} and will not be repeated
here. This ensures that our stellar hydrodynamics part is expected to
perform well on problems involving gravitational contraction/expansion
with and without rotation.

\section{Testing the chemical evolution routines}

Nowadays it is common to find hydrodynamical codes coupled with
passive scalars (or other numerical techniques), able to follow the
evolution of the metals in a simulation.  However, usually no test is
performed to check the numerical accuracy of these routines.
Actually, since a very vast literature on analytical solutions of the
chemical evolution of galaxies and other astronomical objects exists
\citep[see e.g.][and references therein]{Tins80, Matt01, HR10} it is
indeed not difficult to benchmark the chemical evolution routines of a
chemodynamical code.

The ``closed box'' model of chemical evolution is based on the
following assumptions: $(i)$ the system is uniform and closed (there
are no inflows or outflows), $(ii)$ the IMF does not change with time,
$(iii)$ the gas is well mixed at any time and $(iv)$ stars more
massive than a certain threshold mass $M_{thr}$ die instantaneously
whereas stars less massive than $M_{thr}$ live forever.  This last
assumption, although very restrictive, is fulfilled if we look at
specific chemical elements, in particular O and the other
$\alpha$-elements, since they are mostly synthesized by Type II SNe
and the (long-living) low- and intermediate-mass stars contribute
negligibly to their production.

In a system made up of only gas and stars (DM does not play any role
here) we define the quantities

\begin{equation}
R=\int_{M_{thr}}^\infty 
\left[M-M_{rem}(M)\right]\phi(M)dM,\label{eq:r}
\end{equation}
\begin{equation}
y_O=\frac{1}{1-R}\int_{M_{thr}}^\infty 
M p_O(M)\phi(M)dM,\label{eq:yo}
\end{equation}
\noindent
where $M_{tot}=M_{gas}+M_{stars}$, $M_{rem}(M)$ is the remnant mass
after a star of mass $M$ has died and $M p_O(M)$ is the oxygen mass
newly synthesized by the star of mass $M$ ($p_O(M)$ is the true yield
of O).  Under the initial condition $M_{gas}f(0)=Const.=M_{tot}$,
$M_{stars}(0)=0$, $X_O(0)=X_0$ it is not difficult to find that
\begin{equation}
X_O = \frac{M_O}{M_{gas}} = X_0+y_O \ln \left (\mu^{-1}\right)
\label{eq:closedbox},
\end{equation}
\noindent
where $\mu=M_{gas}/M_{tot}$.  This is the famous analytical solution
of the closed box model, which depends only on the gas fraction $\mu$.
  
\begin{figure}
  \resizebox{\hsize}{!}{\includegraphics{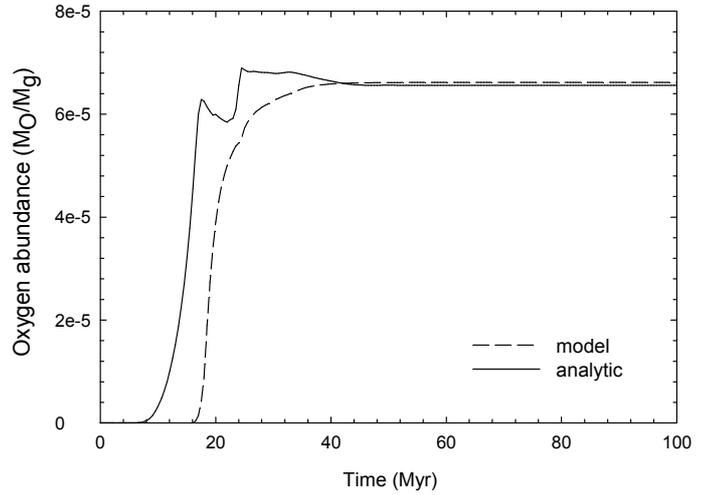}}
  \caption{The evolution with time of the mass fraction of oxygen as
    derived from the analytical expression Eq. \ref{eq:closedbox}
    (solid line) and from our numerical simulation (dashed line).  See
    text for more detail.}
         \label{fig9ap}
\end{figure}

We have integrated numerically eqs. \ref{eq:r} and \ref{eq:yo} by
using the values of $M_{rem}(M)$ and $M p_O(M)$ tabulated by
\citet{WW}, using Z=0.01 Z$_\odot$ as initial metallicity.  We do not
use Z=0 as initial metallicity for two reasons: firstly, for Z(0)=0
the cooling is very low and the resulting star formation is too weak;
secondly, the yields $p_O(M)$ tabulated by \citet{WW} at this
metallicity oscillate considerably and that makes the calculation of
$y_O$ by means of Eq. \ref{eq:yo} less accurate.  By properly choosing
a reference volume in the computational box (namely a volume
encompassing all the metals produced and advected within a time span
of 100 Myr) we can thus calculate the variation of $\mu$ with time
and, from it, the analytical evolution of the oxygen mass fraction
$X_O$ with time.  This is plotted in Fig. \ref{fig9ap} (solid line),
together with the direct numerical calculation of $M_O/M_{gas}$,
averaged over the same volume (dashed line).  The two curves converge
only after $\sim$ 30 Myr.  This is expected because we have to ensure
that Type II SNe of all initial masses contribute to the production of
O and $\sim$ 30 Myr is approximately the lifetime of the less massive
SNeII.  After a time of $\sim$ 40 Myr the two curves almost overlap,
demonstrating that our chemical routines are able to accurately follow
the overall chemical evolution of a galaxy.
  
\end{document}